\documentclass[fleqn,usenatbib]{mnras}

\usepackage{newtxtext,newtxmath}
\usepackage[T1]{fontenc}

\DeclareRobustCommand{\VAN}[3]{#2}
\let\VANthebibliography\thebibliography
\def\thebibliography{\DeclareRobustCommand{\VAN}[3]{##3}\VANthebibliography}

\usepackage{graphicx}	% Including figure files
\usepackage{amsmath}	% Advanced maths commands
\usepackage{gensymb}    % Extra symbol, such as \degree
\usepackage[dvipsnames]{xcolor} % For more colours
\usepackage{ulem}               % \sout command to strike through

\mathchardef\shorthyphen="2D                                            % Short hyphen for system names
\newcommand{\mSun}{\; {\rm M_\odot}}                            % Solar mass
\newcommand{\mJup}{\; {\rm M_{Jup}}}                            % Jupiter mass
\newcommand{\mNep}{\; {\rm M_{Neptune}}}                            % Neptune mass
\newcommand{\mEarth}{\; {\rm M_\oplus}}                         % Earth mass
\newcommand{\mPluto}{\; {\rm M_{Pluto}}}                         % Earth mass
\newcommand{\au}{\; {\rm au}}                                           % Astronomical unit
\newcommand{\m}{\; {\rm m}}                                                     % Metre
\newcommand{\cm}{\; {\rm cm}}                                                     % Centimetre
\newcommand{\myr}{\; {\rm Myr}}                                         % Megayear
\newcommand{\gyr}{\; {\rm Gyr}}                                         % Gegayear
\newcommand{\mPerS}{\; {\rm m \; s^{-1}}}                       % Metres per second
\newcommand{\percent}{\; {\rm per \; cent}}                     % Percent
\newcommand{\rad}{\; {\rm radians} \;}                                         % Radians
\newcommand{\HR}{{\rm HR} \;}                                           % HR number
\newcommand{\mStar}{m_*}
\newcommand{\rIn}{r_{\rm inner}}
\newcommand{\rOut}{r_{\rm outer}}
\newcommand{\aOut}{a_{\rm outer}}
\newcommand{\nDeb}{n_{\rm deb}}
\newcommand{\aDeb}{a_{\rm deb}}
\newcommand{\eDeb}{e_{\rm deb}}
\newcommand{\iDeb}{i_{\rm deb}}
\newcommand{\mPlt}{m_{\rm plt}}
\newcommand{\mPltOverMStar}{\mPlt/m_*}
\newcommand{\aPlt}{a_{\rm plt}}
\newcommand{\ePlt}{e_{\rm plt}}
\newcommand{\aProj}{a_{\rm proj}}
\newcommand{\mProj}{m_{\rm proj}}
\newcommand{\mProjTot}{m_{\rm proj, tot}}
\newcommand{\nProj}{n_{\rm proj}}
\newcommand{\rHill}{r_{\rm Hill, plt}}
\newcommand{\tDiff}{t_{\rm diff}}
\newcommand{\tSec}{t_{\rm sec}}
\newcommand{\tSecProj}{t_{\rm sec,proj}}

\newcommand{\tPlt}{T_{\rm plt}}
\newcommand{\eForced}{e_{\rm forced}}
\newcommand{\QDStar}{Q_{\rm D}^*}

\title[Projectile and resonant stirring]{Increasing planet-stirring efficiency of debris disks by ``projectile stirring'' and ``resonant stirring''}

\author[T. Costa et al.]{
Tyson Costa$^1$\thanks{E-mail: tysoncosta10@gmail.com},
Tim D. Pearce$^{2,1}$\thanks{E-mail: tim.pearce@warwick.ac.uk},
Alexander V. Krivov$^1$
\\
$^1$Astrophysikalisches Institut und Universit\"atssternwarte, Friedrich-Schiller-Universit\"at Jena, Schillerg\"asschen 2-3, D-07745 Jena, Germany\\
$^2$Department of Physics, University of Warwick, Gibbet Hill Road, Coventry CV4 7AL, UK\\
}

\date{Accepted XXX. Received YYY; in original form ZZZ}

\pubyear{2023}

\begin{document}
\label{firstpage}
\pagerange{\pageref{firstpage}--\pageref{lastpage}}
\maketitle

%#############################################################################################################################
\begin{abstract}
Extrasolar debris disks are detected by observing dust, which is thought to be released during planetesimal collisions. This implies that planetesimals are dynamically excited (``stirred''), such that collisions are sufficiently common and violent. The most frequently considered stirring mechanisms are self-stirring by disk self-gravity, and planet-stirring via secular interactions. However, these models face problems when considering disk mass, self-gravity, and planet eccentricity, leading to the possibility that other, unexplored mechanisms instead stir debris. We hypothesize that planet-stirring could be more efficient than the traditional secular model implies, due to two additional mechanisms. First, a planet at the inner edge of a debris disk can scatter massive bodies onto eccentric, disk-crossing orbits, which then excite debris (``projectile stirring''). Second, a planet can stir debris over a wide region via broad mean-motion resonances, both at and between nominal resonance locations (``resonant stirring''). Both mechanisms can be effective even for low-eccentricity planets, unlike secular-planet-stirring.   We run N-body simulations across a broad parameter space, to determine the viability of these new stirring mechanisms. We quantify stirring levels using a bespoke program for assessing {\sc rebound} debris simulations, which we make publicly available. We find that even low-mass projectiles can stir disks, and verify this with a simple analytic criterion. We also show that resonant stirring is effective for planets above ${\sim0.5\mJup}$. By proving that these mechanisms can increase planet-stirring efficiency, we demonstrate that planets could still be stirring debris disks even in cases where conventional (secular) planet-stirring is insufficient.
\end{abstract}

\begin{keywords}
planet-disc interactions -- planets and satellites: dynamical evolution and stability -- circumstellar matter
\end{keywords}

%#############################################################################################################################
\section{Introduction}
\label{sec: introduction}

Debris disks are belts of particles, from planetesimals to small dust grains, surrounding many main-sequence stars (e.g. \citealt{wyatt2008, krivov2010review, matthews2014review, hughes2018review}).  Our own Solar System has two debris disks, the Asteroid Belt and Kuiper Belt, with the latter located beyond Neptune's orbit.

Extrasolar debris disks are detected through observations of micrometre- to millimetre-sized dust grains, in either thermal emission (e.g. \citealt{aumann1984thermal}) or scattered light (e.g. \citealt{smith1984scattered}).  However, where does this dust come from? A dust grain in a debris disk has a relatively short lifetime compared to the host star, before being removed or destroyed by various processes. Dust therefore needs to be continuously replenished from some source.  It is thought that most dust originates from larger planetesimals with longer lifetimes, which collide with each other (e.g. \citealt{weissman1984,harper1984,backman1993}).  This manifests itself as a collisional cascade, where larger bodies in a debris disk collide and break apart into smaller pieces, which in turn collide and break into even smaller pieces, and so on (e.g. \citealt{wyatt1999,wyatt2002}).  For debris particles to reach sufficiently high velocities to break apart in collisions, their eccentricities and/or inclinations must somehow be excited from the almost circular, unstirred orbits they are assumed to have formed with; this excitation process is known as ``stirring'' (\citealt{wyatt2008}).

There are several commonly considered mechanisms for how debris disks get stirred.  The first, secular-planet-stirring, is where a nearby, eccentric planet excites debris through secular interactions \citep{mustill2009}. The efficiency of this process depends on the planet mass, eccentricity, and semimajor axis, but is only possible if the planet's orbit is eccentric.

The most commonly considered alternative mechanism is self-stirring, where larger planetesimals within the disk excite smaller debris (e.g. \citealt{kenyonbromley2001, kenyon2008, kenyonbromley2010selfstirring, kennedy2010, krivov2018}). This process does not require planets to be present. \cite{kenyonbromley2001,kenyonbromley2010selfstirring} proposed that large debris bodies of size $\sim$500-1000~km would be sufficient to stir debris disks.  Later, \cite{krivov2018} concluded that such large objects are not necessary for stirring (they take a long time to form, even though they would stir the disk quickly), and that bodies of $\sim$200~km in size, which form much quicker, are sufficient to self-stir a debris disk.

However, these stirring models do not work in all cases.  For secular-planet-stirring, not only must the planet's orbit be sufficiently eccentric, but the self-gravity of the disk can significantly resist or even cancel the stirring by the planet (L{\"o}hne et al., in prep.).  For self-stirring, the required disk mass is in some cases unphysically high; \cite{krivov2021} concluded that if all debris disks contained objects up to 200~km in size, the masses of some disks would be in the range of ${10^3 - 10^4 \mEarth}$, which exceeds the plausible maximum of ${10^2 - 10^3 \mEarth}$thought to be inheritable from protoplanetary disks. Similarly,  \cite{pearceispy2022} analyzed 67 resolved debris disks and concluded that 26 of these would need masses greater than ${10^2\mEarth}$ to self-stir if considering a conventional debris-size distribution, with 6 of these requiring masses greater than ${10^3 \mEarth}$.  Therefore, in some cases self-stirring appears to require unphysically high debris-disk masses to operate (though also see \citealt{najita2022}).

Several other stirring mechanisms have also been proposed, including ``pre-stirring'' (where debris is already stirred during the protoplanetary disk phase; e.g. \citealt{walmswell2013,boothclarke2016,wyatt2020}) and ``flyby stirring'' (where another star passes close to the system and stirs the debris disk; e.g. \citealt{idalarwood2000,kobayashi2001,kenyonbromley2002}). However, the prevalence of these processes is unclear.

The aim of our study is to look at other possibilities of stirring, which could help explain some cases where secular-planet or self-stirring falter.  We identify two new mechanisms that can significantly increase the efficiency of planet-stirring: ``projectile stirring'' and ``resonant stirring''.  We define projectile stirring to be where a planet scatters one or more large bodies (``projectiles'') from an initially unstirred debris disk onto eccentric orbits.  These newly eccentric projectiles can in turn sufficiently increase the eccentricities of debris within the disk through secular effects, before the planet eventually ejects the projectiles from the system. We will demonstrate that even low-mass projectiles can significantly increase planet-stirring efficiency.

We define resonant stirring to be where a planet stirs a wide region of a debris disk via its mean-motion resonances (MMRs). This effect has historically been overlooked as a stirring mechanism; whilst it is well known that MMRs can excite debris in narrow regions around nominal MMR locations (e.g. \citealt{Dermott1983, Krivov2007, Tabeshian2016, Malhotra2019}), it is often overlooked that individual MMRs can be very broad, particularly for high-mass planets (e.g. \citealt{murray1999}). This means that MMRs can significantly excite debris not only at the nominal resonance locations, but \textit{between} those locations as well. We will demonstrate that broad MMRs can stir debris across a very wide region of a debris disk, far wider than simply around the nominal MMR locations as might na\"{\i}vely be assumed. We will also show that this effect is efficient even for planets with very low eccentricities, in contrast to the traditional, secular model of planet-stirring.

The concept of projectile stirring is motivated by the increasing prevalence of suspected scattered populations in debris disks. The Kuiper Belt hosts a scattered population, comprising bodies that previously had close encounters with Neptune \citep{Gomes2008}. This population is highly eccentric, and these scattered bodies cross the orbits of other, dynamically colder Kuiper-Belt Objects. The scattered population includes dwarf planets such as Eris, demonstrating that high-mass bodies can be scattered onto disk-crossing orbits by planets located at the inner edges of debris disks. There are also tentative indications that even larger bodies may once have existed in this region, which got scattered and subsequently ejected from the Solar System (e.g. \citealt{Huang2022}). These arguments motivate questions about the dynamical impact of scattered bodies on the excitation level of debris disks. In addition to the Solar System, there is also evidence of scattered material in extrasolar debris disks; \cite{geiler2019} found evidence for a scattered disk in ${\HR8799}$, and \cite{matra2019} identified both ``hot'' and ``cold''  dynamical populations in the \mbox{$\beta$ Pictoris} debris disk. In addition, the sharp inner edges of some debris disks could indicate the presence of sculpting planets at those locations \citep{ImazBlanco2023, Pearce2023}, and such planets would generate scattered populations. Based on both the Solar System and extrasolar systems, scattered populations could be a common feature of debris disks; given that observed disks must be dynamically excited, and that there are problems associated with known stirring mechanisms, it is pertinent to ask whether scattered material could be helping to stir observed debris disks.

Both projectile and resonant stirring are new ways to increase planet-stirring efficiency beyond the original secular effect of \cite{mustill2009}.  \cite{munozmixedstirring2023} have also recently investigated a similar topic that they refer to as ``mixed stirring''. In their work, they simulated systems with both a giant planet and a debris disk comprising massive bodies, in order to investigate stirring through a combination of effects (see also \citealt{munoz2015dpdiskexcitation}).  In contrast, we focus on two specific stirring processes (projectile and resonant stirring) and isolate them from other interactions to assess their individual effectiveness.  These effects likely also occur in the simulations of \cite{munozmixedstirring2023}.

In this paper we run $N$-body simulations across a wide parameter space, to quantify the effectiveness of projectile and resonant stirring. We first describe how we quantify stirring, including providing a public code to quantify the stirring level in a {\sc rebound} simulation (\mbox{Section \ref{sec: stirring_analysis}}). We then present the setup and results of our N-body simulations (\mbox{Section \ref{sec: simulations}}). In \mbox{Section \ref{sec: projectileStirring}} we investigate projectile stirring in more detail, including deriving an analytic criterion for when it will occur (\mbox{Equation \ref{eq: theoMinProjMassToStir}}), and in \mbox{Section \ref{sec: resonant_stirring}} we further investigate resonant stirring. In \mbox{Section \ref{sec: discussion}} we discus our results, and we conclude in Section \ref{sec: conclusion}.  The Appendix contains more details about our stirring analyses.

%#############################################################################################################################
\section{Stirring definitions}
\label{sec: stirring_analysis}

In Section \ref{sec: simulations}, we will run N-body simulations to investigate the effectiveness of projectile and resonant stirring. However, before proceeding we must first define quantitatively what we mean by stirring. For this paper we consider two different stirring definitions; first, a simple analytic form that quantifies the minimum eccentricity that two adjacent bodies would need to collide destructively, and second, a detailed numerical approach that properly accounts for the specific orbits and orientations. The former provides a rough idea of whether a disk is stirred or not, whilst the latter provides a much more accurate picture and is the method that we use to assess our simulations. We outline both analyses in this section, and give more details in Appendix \ref{sec: stirring_analysis_appendix}.

%-----------------------------------------------------------------------------------------------------------------------------
\subsection{Rough criterion for debris to be considered stirred}
\label{subsec: simpleStirringCriterion}

We first perform a simple, rough calculation of the minimum eccentricity that a debris particle must attain to be considered stirred. We consider a system containing a star and two massless debris bodies. We assume that these particles have co-planar orbits, and that the two orbits are rotationally offset but otherwise identical (see Figure \ref{fig: simple_velocity_diagram}). In Appendix \ref{sec: simplified_stirring_analysis_appendix} we show that such bodies are unstirred if their eccentricities are below some value:

\begin{equation}
    e_{\rm unstirred} < \frac{v_{\rm frag}}{2}\sqrt{\frac{\aDeb}{G \mStar}},
    \label{eq: unstirred_ecc}
\end{equation}

\noindent where $v_{\rm frag}$ is the fragmentation speed of a debris body, which is dependent on its size and material, $G$ is the gravitational constant, $\aDeb$ is the semimajor axis of the debris particles, and $\mStar$ is the star mass.  We can also relate this to $v_{\rm circ}$, the speed of a circular orbit at the debris semimajor axis; since ${v_{\rm circ}^2 = G\mStar/\aDeb}$, Equation \ref{eq: unstirred_ecc} can alternatively be written as

\begin{equation}
    e_{\rm unstirred} < \frac{v_{\rm frag}}{2v_{\rm circ}}.
    \label{eq: unstirred_ecc_v_circ}
\end{equation}

For this paper we want to evaluate Equation \ref{eq: unstirred_ecc}, to roughly constrain whether debris could be stirred as a function of its semimajor axis. This requires us to evaluate the fragmentation speed, $v_{\rm frag}$, which means we must make assumptions about debris size and composition. First, in line with previous studies, we approximate the material to be basalt using the prescription of \citet{wyatt2002}. This yields a size-dependent form for $v_{\rm frag}$, given by Equation \ref{eq: fragmentation_velocity_app}. Then, we assume the debris particles to have radii of ${1\cm}$; this is because debris-disk observations are sensitive to grains of millimetre size or smaller, so larger particles must be undergoing destructive collisions to produce the observed dust. Hence a pair of centimetre-sized, basalt grains on identical, azimuthally rotated orbits are unstirred if their eccentricity is less than

\begin{equation}
    e_{\rm unstirred}(1~{\rm cm}) < 8.9 \times 10^{-4} \left( \frac{\aDeb}{\rm au}\right)^{1/2} \left( \frac{\mStar}{\mSun}\right)^{-1/2}.
    \label{eq: final_unstirred_ecc}
\end{equation}

\noindent Note that having an eccentricity above this value is not a sufficient condition to be stirred, since collisional speed depends on relative orientation to adjacent-particle orbits; nonetheless, debris whose eccentricities never exceed this critical value are almost certainly unstirred. The Equation \ref{eq: final_unstirred_ecc} criterion will be displayed on figures throughout our paper, to provide a rough idea of the stirring level in N-body simulations.

%-----------------------------------------------------------------------------------------------------------------------------
\subsection{Full, numerical stirring analysis}
\label{subsec: numericalStirringAnalysis}

Equation \ref{eq: final_unstirred_ecc} is only a rough estimate of whether debris is stirred, because it assumes that nearby debris are on identical, optimally misaligned orbits. A real disk may contain debris with a wide range of orbits and orientations. To properly assess stirring in our N-body simulations, we will therefore use a numerical method instead.

The principle of our numerical stirring analysis is to take the results of an N-body simulation, and post-process it to determine the stirring level. We do this by numerically analysing pairs of debris particles with instantaneously overlapping orbits, to calculate the collision speed at the orbit-intersection points. If this speed is greater than the fragmentation speed, again assuming ${1\cm}$ basalt grains, then both particles are considered stirred and are not included in future stirring calculations. Conversely, if the collision speed is below the fragmentation speed, then a different pair of particles is considered instead. Only pairs of particles that were initially near each other are considered, so the stirring level at the outer edge of a broad disk would not be contaminated by high-eccentricity debris scattered by a planet at the inner edge. The process continues until all suitable particle pairs have been checked, and is then repeated across multiple simulation snapshots at different times. This yields all debris particles that have ever been stirred in a simulation, and is much more accurate at assessing the disk's stirring level than Equation \ref{eq: final_unstirred_ecc}, because it properly accounts for different orbit orientations and semimajor axes. The full details of this numerical analysis are given in Appendix \ref{sec: general_stirring_analysis_appendix}, and we publicly release the code as a {\sc python} programme to quantify stirring in {\sc rebound} N-body simulations\footnote{\href{https://www.tdpearce.uk/public-code/}{tdpearce.uk/public-code}}. We will use this numerical analysis to assess stirring in all simulations in this paper.

%#############################################################################################################################
\section{N-body simulations}
\label{sec: simulations}

To assess the effectiveness of projectile and resonant stirring, we run a large suite of N-body simulations, covering a wide range of system setups. Individual setups are randomly generated, but have similar qualitative characteristics as described in the following section; each comprises at least a star, a planet, and a massless, external debris disk. For each setup, we run two simulations; the first comprises only the star, planet, and disk, whilst the second also contains several massive debris bodies (projectiles). This approach lets us distinguish the stirring effects of the planet and projectiles. Once each simulation has finished, we apply our numerical analysis code (Section \ref{subsec: numericalStirringAnalysis}) to quantify the stirring level. In Section \ref{subsec: multi_simulation_setup} we describe the setups of our N-body simulations, in Section \ref{subsec: simulationStrategy} our simulation strategy, and in Section \ref{subsec: multi_sim_results} we present the results of an example pair of simulations.

%-----------------------------------------------------------------------------------------------------------------------------
\subsection{Simulation setup}
\label{subsec: multi_simulation_setup}

\begin{figure}
    % discPlanetInteractions/arks/discInnerEdges/report/imageConstruction/setupDiagram.vsz
    \includegraphics[width=8cm]{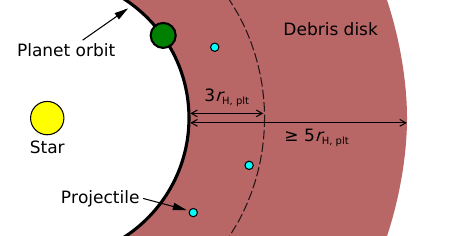}
    \caption{Initial setup of our N-body simulations. A single planet is initialised on a near-circular orbit around a star. A massless debris disk is initialised with its inner edge at the planet's orbit, and its outer edge at least 5 Hill radii beyond this (in most cases, the disk extends well beyond 5 Hill radii). Half of our simulations also include massive debris bodies (``projectiles''); if present, these are initialised on low-eccentricity orbits within 3 Hill radii of the planet's orbit, to ensure that they are scattered by the planet. The ranges of our specific simulation parameters are listed in Table \ref{tab: random_sim_setup}.}
    \label{fig: setupDiagram}
\end{figure}

Our aim is to investigate a scenario where a planet scatters the inner region of an external debris disk, whilst leaving the outer region unscattered. By naturally generating a scattered population, we can investigate the dynamical effect of scattered bodies (projectiles) on the rest of the disk. Having a sufficiently wide disk also lets us investigate the effect of the planet on distant debris, well beyond the scattering region. To this end, we set up our simulations with the debris disk initially extending from the planet's orbit out to at least \mbox{5 Hill} radii beyond this. We provide a setup diagram on Figure \ref{fig: setupDiagram}, and describe the setup and parameters in this section.

%-----------------------------------------------------------------------------------------------------------------------------
\subsubsection{Integrator}

Our N-body simulations are performed using the \textsc{python} version of \textsc{rebound} \citep{rebound}, with the \textsc{mercurius} hybrid-symplectic integrator \citep{reboundmercurius}. This uses a symplectic Wisdom-Holman integrator ({\sc whfast}) when particles are far apart, and switches to a high-order integrator ({\sc ias15}) during close encounters. This lets us efficiently and accurately model both individual scattering events and long-term evolution.

%-----------------------------------------------------------------------------------------------------------------------------
\subsubsection{General setup}
\label{subsec: planetOnlySimSetup}

Our randomised system setups are designed to cover a wide parameter space, but also be roughly representative of real debris-disk systems (e.g. those in Table 1 of \citealt{matra2018}). Our parameter space is listed in Table \ref{tab: random_sim_setup}. When initialising a system, we first draw the star mass $\mStar$ from the range ${0.08 \leq \mStar \leq 2\mSun}$, using a uniform-logarithmic distribution. We then draw the planet mass $\mPlt$ from the range ${1\mNep \leq \mPlt < 10\mJup}$ (equivalent to ${17 \leq \mPlt/{\rm M_\oplus} < 3200}$), again using a uniform-logarithmic distribution. The planet's semimajor axis is uniformly drawn between 12.5 and ${100\au}$, and its eccentricity\footnote{A non-zero eccentricity is required to ensure proper behaviour in \textsc{rebound}, but the planet's initial eccentricity is always much smaller than that required to secularly stir the disk.} set to ${10^{-4}}$.

\begin{table*}
    \centering
    \caption{Parameters, distribution types, and values or ranges used to set up the randomized debris-disk simulations. Arrows specify ranges from a lower value (left) to a higher value (right).}
    \label{tab: random_sim_setup}
    \begin{tabular}{lccc} % two columns, alignment for each
	  \hline
	  Parameter name & Symbol & Distribution type & Value or range\\
        \hline
        Star mass & $\mStar$ & Uniform logarithmic & $0.08 \rightarrow 2~{\mSun}$\\
        Planet mass & $\mPlt$ & Uniform logarithmic & ${1\mNep \; (17\mEarth) \rightarrow 10\mJup \; (3200\mEarth)}$\\
        Planet semimajor axis & $\aPlt$ & Uniform & $12.5 \rightarrow 100\au$\\
        Planet eccentricity & $\ePlt$ & - & $10^{\rm -4}$\\
        Disk inner edge & $\rIn$ & - & $\aPlt$\\
        Disk outer edge & $\rOut$ & \multicolumn{2}{c}{See calculations in Section \ref{subsec: planetOnlySimSetup}}\\
        Disk surface-density profile & - & - & $\propto \aDeb^{-{3/2}}$ \\
        Debris eccentricities & $\eDeb$ & Uniform & $0 \rightarrow 10^{-4}$ \\
        Debris inclinations & $\iDeb$ & Uniform & $0 \rightarrow 10^{-4}\rad$ \\   
        Number of massless debris particles & $\nDeb$ & - & 2000\\
        Mass of individual projectiles & $\mProj$ & Uniform logarithmic & ${1\mPluto \; (2.2\times10^{-3}\mEarth) \rightarrow 0.1\mPlt}$\\
        Number of projectiles & $\nProj$ & - & 0 or 10 \\
        \hline
    \end{tabular}
\end{table*}

The massless debris disk is initialised with its inner edge, $\rIn$, equal to the planet's semimajor axis $\aPlt$. The disk's outer edge is drawn to be at least \mbox{5 Hill} radii ($5\rHill$) beyond this, to ensure that some of the disk remains unscattered by the planet\footnote{A planet will typically scatter non-resonant material originating within a little over \mbox{3 Hill} radii of its orbit \citep{gladman1993, ida2000, kirsh2009, Malhotra2021, friebe2022, Pearce2023}.}. To avoid a bias towards generating wide disks, the outer-edge assignment is performed by first drawing the disk's width-to-radius ratio, ${\Delta R / R}$. This is uniformly drawn in the range ${(\Delta R / R)_{\rm min}}$ to 1, where ${(\Delta R / R)_{\rm min}}$ is

\begin{equation}
    \left(\frac{\Delta R}{R}\right)_{\rm min} = 2 \times \frac{\aPlt + 5\rHill - \rIn}{\aPlt + 5\rHill + \rIn}.  
    \label{eq: width_ratio}
\end{equation}

\noindent Once $\Delta R/R$ is drawn, the disk's outer edge, $\rOut$, can be calculated as

\begin{equation}
    \rOut = \rIn \times \frac{2 + \Delta R/R}{2 - \Delta R/R}.
    \label{eq: outer_disk_radius}
\end{equation}

The debris disk comprises 2000 massless debris particles. Each particle has its semimajor axis $\aDeb$ drawn such that the disk surface-density profile goes as $\aDeb^{-3/2}$, like the Minimum-Mass Solar Nebula (\citealt{Weidenschilling1977, Hayashi1981}). Each particle has an individual eccentricity $\eDeb$ uniformly drawn in the range ${0 \leq \eDeb < 10^{-4}}$, and an inclination $\iDeb$ relative to the planet's orbital plane uniformly drawn in the range ${0 \leq \iDeb \leq 10^{-4} {\rm \; radians}}$. Each particle's longitude of ascending node, argument of pericentre, and mean anomaly are individually drawn from uniform distributions between $0$ and ${2\pi {\rm \; radians}}$.

The above setup describes simulations without additional, massive projectiles, such that the planet is the only body perturbing the debris disk. For these simulations, the setup is almost complete.  We set the {\sc whfast} fixed timestep to be ${1\percent}$ of the planet's period, and the minimum value of the {\sc ias15} variable timestep to be ${10^{-6}}$ times the planet's period. During the integration, any body is removed from the simulation if it reaches a threshold distance of ${10^4\au}$.

We run each simulation until 10 diffusion timescales is reached, where the diffusion timescale $\tDiff$ characterises the scattering timescale. This ensures that scattering is essentially complete by the end of the simulation, as we will later demonstrate in Section \ref{subsec: particle_scattering}. Formally, one diffusion timescale quantifies how long it takes for the energy of a particle undergoing scattering to change by order of itself, and is given by

\begin{equation}
    \tDiff \approx 0.01 \tPlt \left(\frac{\aPlt}{a}\right)^{\frac{1}{2}} \left(\frac{\mPlt}{\mStar}\right)^{\rm -2},
    \label{eq: diffusion_time}
\end{equation}

\noindent where $\tPlt$ is the planet's orbital period, and $a$ is the semimajor axis of the scattered body (\citealt{tremaine1993}; Equation 18 in \citealt{Pearce2014}). To determine the end time of our simulations, we evaluate $\tDiff$ at the planet's location, i.e. at ${a=\aPlt}$.

%-----------------------------------------------------------------------------------------------------------------------------
\subsubsection{Projectile simulations}
\label{subsec: projectileSimSetup}

Half of our simulations are set up as in Section \ref{subsec: planetOnlySimSetup}. For the other half, we also include massive debris bodies (projectiles). These bodies typically represent dwarf planets, and are initialised such that they are very likely to get scattered onto high-eccentricity orbits by the planet. In the projectile simulations, the star, planet, massless disk, and integrator parameters are again set up exactly as in Section \ref{subsec: planetOnlySimSetup}. We then initialise 10 projectiles, each with the same mass $\mProj$, where for each simulation $\mProj$ is drawn from a uniform-logarithmic distribution between a Pluto mass ({$2.2\times10^{-3}\mEarth$}) and ${10\percent}$ of the planet's mass. Each projectile's semimajor axis $\aProj$ is individually drawn between the planet's semimajor axis and \mbox{3 planetary} Hill radii beyond this, such that the projectiles are initially located close enough to the planet to undergo scattering. Like massless debris, the projectile's semimajor axes are drawn from a $\aProj^{-3/2}$ profile, and their individual eccentricities and inclinations are uniformly drawn between 0 to ${10^{-4}}$ and 0 to ${10^{-4} {\rm \; radians}}$ respectively. Also, each projectile's longitude of ascending node, argument of pericentre, and mean anomaly are individually drawn from uniform distributions between $0$ and ${2\pi {\rm \; radians}}$. Simulations are again run to 10 diffusion timescales, as calculated from the initial orbit of the planet.

%-----------------------------------------------------------------------------------------------------------------------------
\subsection{Simulation strategy}
\label{subsec: simulationStrategy}

For each system setup we run two simulations; one with projectiles, and one without. Both simulations use the same seed, such that the star, planet, and massless disk are identical between the two. We run 160 pairs of randomized simulations. We omit any simulations where projectiles significantly excite the planet eccentricity (above 0.01), because this results in the eccentric planet secularly stirring the disk, which is not the scenario we investigate.

After each N-body simulation finishes, we run the numerical stirring analysis from Section \ref{subsec: numericalStirringAnalysis} to assess the state of the disk. The code classes each debris particle as either ``scattered'' (i.e. it was ejected, or its semimajor axis changed by at least ${20\percent}$), or ``stirred'' or ``unstirred'' (from the assessment described in Section \ref{subsec: numericalStirringAnalysis} and Appendix \ref{sec: general_stirring_analysis_appendix}). We bin the debris particles by their initial semimajor axis, to assess the degree of scattering and stirring at multiple locations across the disk.

%-----------------------------------------------------------------------------------------------------------------------------
\subsection{Example simulation results}
\label{subsec: multi_sim_results}

\begin{figure*}
    % projectileAndResonantStirring/images/nBodyElements/nBodyElements.vsz
    % This is Sim 131
    \includegraphics[width=12.6cm]{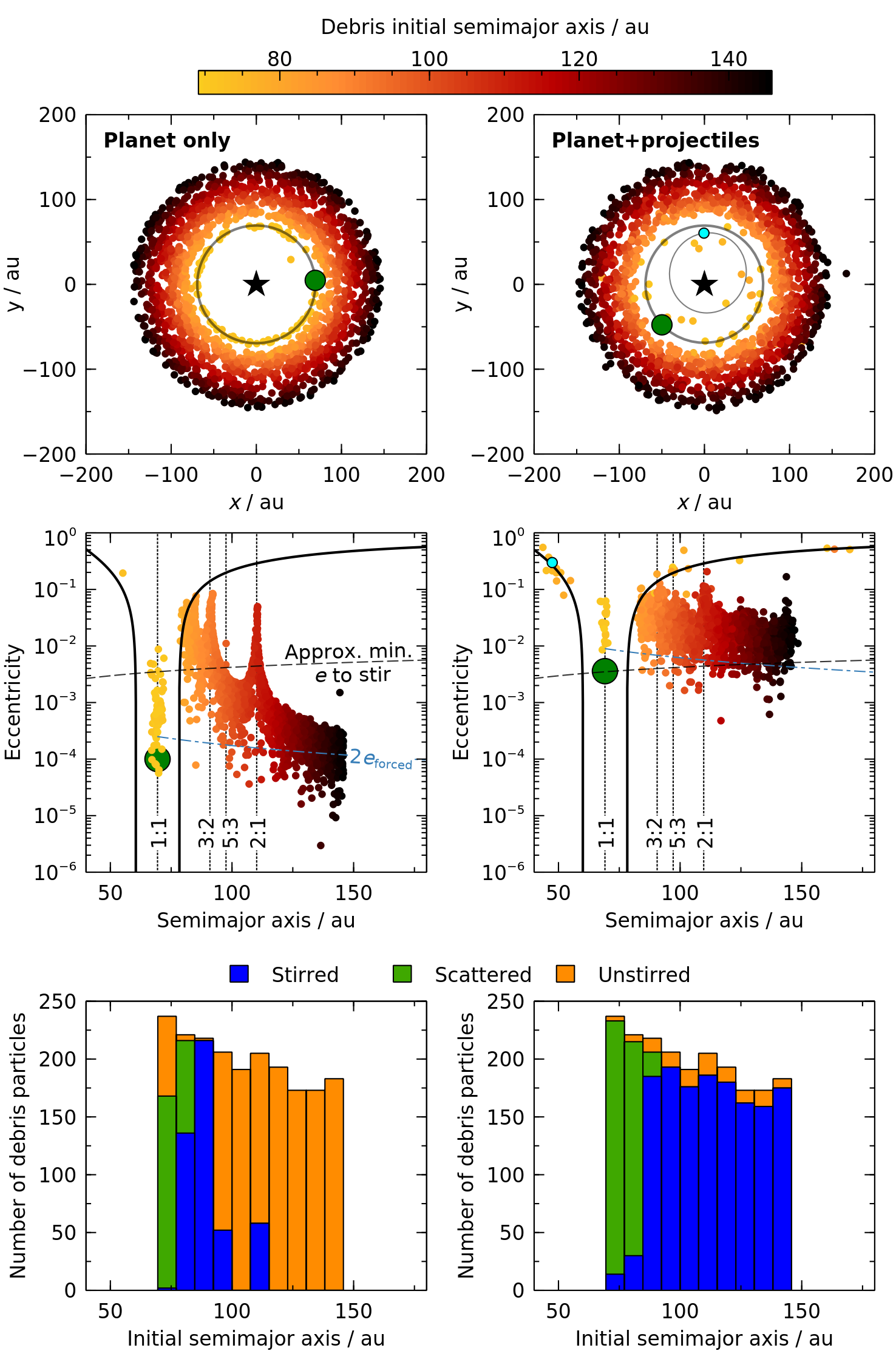}
    \caption{Comparison of a planet-only \textsc{rebound} simulation (left) to an equivalent simulation that also includes 10 massive projectiles (right). The simulations comprise a ${1.7\mSun}$ star, a ${0.41\mJup}$ planet at ${69\au}$, and a massless debris disk extending to ${146\au}$; in the projectile simulation, each projectile has a mass of ${0.088\mEarth}$. The simulations are shown after ${83\myr}$ (ten diffusion timescales). Top: Systems in the \textit{x}-\textit{y} plane.  The large green circle is the planet, the thick line its orbit, with the star at the origin.  Debris particles are the smaller circles, coloured by their initial semimajor axes. Only one projectile survives until the end of the simulation; this is the cyan circle, on an orbit shown by the thin line. Middle: Eccentricities and semimajor axes; the vertical dotted lines are nominal mean-motion resonance locations, the dashed line is an approximation of the minimum eccentricity required for $1\cm$ debris to be stirred (Equation \ref{eq: final_unstirred_ecc}), the solid black lines show where particles come within 3 Hill radii of the planet's orbit, and the dot-dash blue line is twice the forcing eccentricity (the maximum eccentricity expected if debris interacts secularly with the planet; Equation \ref{eq: forced_ecc}).  Bottom: Number of debris particles stirred, scattered, and unstirred in the two simulations, from the full, numerical stirring analysis (Appendix \ref{sec: general_stirring_analysis_appendix}).  As seen in the right plots, including massive projectiles scattered by a planet can significantly increase the stirring level in the disk. The planet-only simulation also shows some degree of resonant stirring, as discussed in Section \ref{sec: resonant_stirring}.}
    \label{fig: projectileVsPltOnlySims}
\end{figure*}

Figure \ref{fig: projectileVsPltOnlySims} shows an example pair of simulations; the left panels show the result of the planet-only simulation, and the right panels the equivalent simulation with the addition of 10 massive projectiles.

In both simulations, the planet clears the majority of debris initially within \mbox{3 Hill} radii of its orbit, with most debris beyond this surviving until the end of the simulation. In the planet-only simulation (left plots), the planet excites debris at the disk inner edge, as well as near the nominal location of several MMRs, but the majority of the surviving disk remains unstirred. Conversely, the addition of projectiles greatly increases the stirring level (right plots); debris eccentricities are excited to a much higher degree, and the surviving disk is almost completely stirred. Nine of the ten projectiles were ejected during the simulation.

The above results are for one specific system setup. Across all of our setups, we generally find that the addition of projectiles increases the excitation level of the disk, but the degree of stirring varies across systems. We also find that some planet-only simulations are able to stir very broad regions through MMRs alone. In Sections \ref{sec: projectileStirring} and \ref{sec: resonant_stirring} we assess projectile and resonant stirring across all of our simulations, including predicting when these mechanisms are important.

%#############################################################################################################################
\section{Projectile stirring}
\label{sec: projectileStirring}

Over half of our projectile simulations show a significant increase in debris-stirring level, like the simulation on the right panels of Figure \ref{fig: projectileVsPltOnlySims}. In this section we analyse the mechanism in greater detail. In Section \ref{subsec: isProjStirSecOrScat} we determine the dynamical interaction by which projectiles stir debris, in Section \ref{subsec: projMassAndNumber} we examine how the number and mass of projectiles affects stirring, and in Section \ref{subsec: analyticPredOfProjectileStirring} we derive an analytic criterion for when we expect projectile stirring to occur, and verify it across our simulations.

%-----------------------------------------------------------------------------------------------------------------------------
\subsection{Do projectiles excite debris via secular or scattering interactions?}
\label{subsec: isProjStirSecOrScat}

We first determine the dynamical effect by which projectiles excite debris eccentricities. There are two plausible interactions: secular or scattering. In this section we analyse both, and demonstrate that secular interactions are the means by which projectiles stir debris. We will do this by comparing the timescales associated with each interaction.

%-----------------------------------------------------------------------------------------------------------------------------
\subsubsection{Secular interactions between debris and projectiles}
\label{subsec: projSecular}

The first dynamical effect we consider is the secular effect of a high-eccentricity, disk-crossing projectile on other debris particles. Secular effects are long-term interactions, occurring on timescales much longer than orbital periods. They cause orbits to periodically oscillate in eccentricity, inclination, and orientation, whilst semimajor axes remain constant. Secular oscillations will be further discussed in Section \ref{subsec: identifyingMMRsAsTheStirrers}, but for this section it is sufficient to note that secular interactions cause debris eccentricities to oscillate.

The period of these oscillations is the secular timescale, $\tSec$:

\begin{equation}
    \tSec \approx \begin{cases}
    4 \tPlt \left( \frac{\mPlt}{\mStar}\right)^{-1} \left(\frac{\aPlt}{\aDeb}\right)^{-\frac{5}{2}} \left[ b_{3/2}^{(1)} \left(\frac{\aPlt}{\aDeb}\right)\right]^{-1}, & \text{if } \aPlt < \aDeb; \vspace{2mm} \\
    4 \tPlt \left( \frac{\mPlt}{\mStar}\right)^{-1} \left(\frac{\aDeb}{\aPlt}\right)^{-\frac{1}{2}} \left[ b_{3/2}^{(1)} \left(\frac{\aDeb}{\aPlt}\right)\right]^{-1}, & \text{else},
  \end{cases}
\label{eq: secular_time}
\end{equation}

\noindent where $b_{\rm 3/2}^{(\rm 1)} (\alpha)$ is a Laplace coefficient:

\begin{equation}
    b_{\rm s}^{(\rm j)} (\alpha) \equiv \frac{1}{\pi} \int_{\rm 0}^{\rm 2\pi} \frac{\cos({j\psi})}{(1 - 2\alpha\cos{\psi} + \alpha^{\rm 2})^{\rm s}} d\psi
    \label{eq: laplace_coeff}
\end{equation}

\noindent \citep{murray1999}. Since the secular time is dependent on particle semimajor axis, debris particles farther away from the massive body will take longer to undergo one full period of the secular interaction.

The above secular framework is based on Laplace–Lagrange theory, which is strictly only valid for low-eccentricity, non-crossing orbits. However, the rough principles still hold when applied to high-eccentricity, crossing orbits \citep{beust2014, Beust2016, Pearce2014, Pearce2015, Pearce2021}. We will therefore use the Laplace–Lagrange secular timescale to quantify the secular effect of high-eccentricity projectiles on debris.

%-----------------------------------------------------------------------------------------------------------------------------
\subsubsection{Scattering of debris by projectiles}
\label{subsec: projScattering}

The second dynamical effect we consider is scattering. Scattering occurs when two bodies make a close approach to each other; in the context of projectile stirring, we investigate whether projectiles on highly eccentric, disk-crossing orbits excite debris through close approaches to individual debris. Scattering is a short-term interaction, occurring on timescales much shorter than the orbital periods.

In Section \ref{subsec: planetOnlySimSetup} we argued that scattering is quantified by the diffusion timescale (Equation \ref{eq: diffusion_time}). Specifically, we argued that this timescale is related to how long it takes a larger body to excite or eject smaller bodies via scattering. Like the secular timescale (\mbox{Equation \ref{eq: secular_time}}), the diffusion timescale was originally derived assuming the massive body to be on a low-eccentricity orbit \citep{tremaine1993}; however, it is reasonably accurate for high-eccentricity orbits as well \citep{Pearce2014, Pearce2015}. We therefore use the diffusion timescale to quantify the scattering of debris by a high-eccentricity projectile.

%-----------------------------------------------------------------------------------------------------------------------------
\subsubsection{Secular or scattering?}
\label{subsec: projSecularVsScattering}

\begin{figure}
    \centering
    \includegraphics[width=8cm]{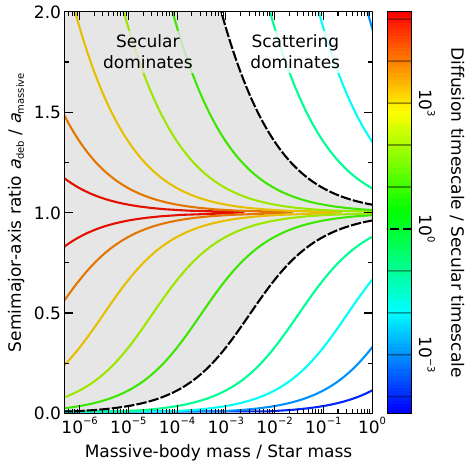}
    \caption{The ratio of secular and diffusion timescales (Equations \ref{eq: diffusion_time} and \ref{eq: secular_time}), which determines whether the interaction between an eccentric, massive body and a test particle will be dominated by secular or scattering effects.  The black dashed line is where the diffusion timescale is equal to the secular timescale; secular interactions will dominate if the secular timescale is shorter than the diffusion timescale (shaded region), and \textit{vice-versa}.  Since our considered projectiles are typically dwarf planets, the plot shows that the projectile's diffusion timescale is almost always much longer than the secular timescale, so the interaction between projectiles and a debris disk will be predominantly secular in nature.}
    \label{fig: dwarf_timescale_comparison}
\end{figure}

To assess whether projectiles would excite debris via secular or scattering interactions, we compare the secular and scattering timescales from Equations \ref{eq: diffusion_time} and \ref{eq: secular_time}. Figure \ref{fig: dwarf_timescale_comparison} shows the ratio of these timescales for an interaction between a massive body and a test particle; for an eccentric, massive body, secular interactions will dominate if the secular timescale is shorter than the diffusion timescale, and \textit{vice-versa}.

Figure \ref{fig: dwarf_timescale_comparison} is general for any interaction between an eccentric, massive body and a test particle. We now apply this to projectile stirring; specifically, to determine the dynamical effect of a projectile (i.e. the massive body on Figure \ref{fig: dwarf_timescale_comparison}) on massless debris particles. The projectiles we consider are typically large planetesimals or dwarf planets, so the mass ratio $\mProj/\mStar$ is always very small ($\lesssim 10^{-3}$). For this regime, Figure \ref{fig: dwarf_timescale_comparison} shows that in almost all cases the diffusion timescale will be longer than the secular timescale, i.e. in the shaded region of the plot.  The conclusion is that, in almost all cases, the secular effect of the projectile will have greater impact on debris than scattering; this means that projectiles stir debris via secular interactions, rather than by scattering.

%-----------------------------------------------------------------------------------------------------------------------------
\subsection{Projectile-stirring efficiency as a function of mass and number of projectiles}
\label{subsec: projMassAndNumber}

In Section \ref{subsec: multi_sim_results} we presented an example pair of simulations, which are good representations of the majority of our simulations. These showed that stirring is significantly increased when projectiles are included. In that example there were 10 equal-mass projectiles, with a total mass of ${0.88\mEarth}$. We now examine how the stirring level changes if we vary the mass and number of projectiles; the quantitative results will be specific to that example system, but they qualitatively hold for all of our simulations.

To better visualize the results of the increase in stirring done by projectiles for a particular system (instead of generalizing across any system), we can also plot the stirring percentage increase against the total-projectile mass to determine a critical mass for when stirring is likely to occur.

To quantify projectile stirring, we first define a metric. This will provide a single number that quantifies how much the stirring level increases due to the presence of projectiles. Since the planet on its own is able to stir some debris through MMRs (\mbox{Section \ref{sec: resonant_stirring}}), we must account for this planet-only stirring when assessing the effect of projectiles. To proceed for a given system, we run our numerical stirring analysis on the planet-only simulation, and then separately on the projectile simulation. We then quantify the degree of projectile stirring as

\begin{equation}
    \begin{aligned}
        &\text{projectile-stirring efficiency} = \\
        &\frac{\text{particles~stirred~in~proj.~sim. - particles~stirred~in~planet~sim.}}{\text{particles~unstirred~in~planet~sim.}},
    \end{aligned}
    \label{eq: stir_percent_increase}
\end{equation}

\noindent i.e. the increase in the number of debris particles stirred in the projectile simulation relative to the planet-only simulation, divided by the number that were not stirred in the planet-only simulation. For the purposes of this evaluation, we also class scattered and ejected particles as ``stirred'', since they have been excited. According to this definition, if projectiles do not affect the stirring outcome then the projectile-stirring efficiency is ${0\percent}$; conversely, if projectiles stir all bodies that were not stirred by the planet alone, then the efficiency is ${100\percent}$. We apply this metric to quantify how the stirring level changes as a function of the mass and number of projectiles in the following sections.

%-----------------------------------------------------------------------------------------------------------------------------
\subsubsection{Projectile mass}
\label{subsec: massOfProjectiles}

To assess the effect of projectile mass on the stirring level, we re-run the projectile simulation from \mbox{Figure \ref{fig: projectileVsPltOnlySims}} with different projectile masses. In each simulation the 10 projectiles always have equal masses. For each projectile mass we test, we run multiple simulations; the planet, star, and disk parameters are kept constant, whilst the initial projectile orbits are randomised.

\begin{figure}
    % image: projectileAndResonantStirring/images/stirringIncreaseByProjectiles/stirringIncreaseByProjectiles.vsz
    % data: projectileAndResonantStirring/images/stirringIncreaseByProjectiles/data/sim110DifferentProjMassesAndNumbers.ods
    % Run projectileAndResonantStirring/runStirringCheckCodeOnAllSimsFixedPars.py and copy results into ods
    \centering
    \includegraphics[width=8cm]{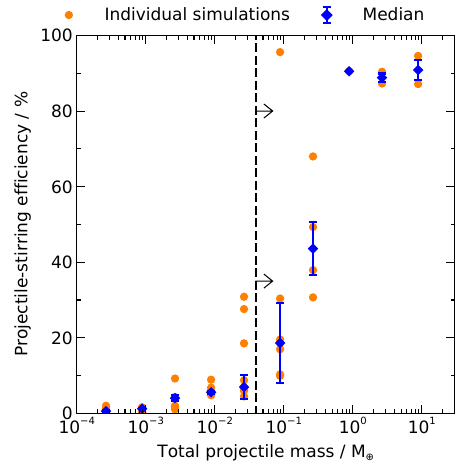}
    \caption{Efficiency of projectile stirring for the example system on Figure \ref{fig: projectileVsPltOnlySims}, as a function of the total-projectile mass. In each case the total-projectile mass is distributed between 10 equal-mass projectiles, and the projectile-stirring efficiency calculated using Equation \ref{eq: stir_percent_increase}. As the total mass of projectiles increase, the amount of stirring also increases. From these simulations, the minimum total-projectile mass required for stirring is ${\sim0.05\mEarth}$. This agrees with the theoretical prediction of ${0.040\mEarth}$ from Equation \ref{eq: theoMinProjMassToStir}, shown by the dashed vertical line.}
    \label{fig: stirringIncreaseVsProjectileMass}
\end{figure}

Figure \ref{fig: stirringIncreaseVsProjectileMass} shows the results, where we quantify the projectile-stirring efficiency using Equation \ref{eq: stir_percent_increase}. As could be expected, the higher the projectile mass, the more efficiently the disk is stirred. For this specific system, projectile stirring has a significant effect if the total-projectile mass is greater than ${\sim0.05\mEarth}$. Later, in Section \ref{subsec: analyticProjStirringTimescaleCriterion}, we will use an analytic argument to predict the minimum total-projectile mass required for projectile stirring to occur; that analysis predicts a minimum total-projectile mass of ${0.040\mEarth}$, which agrees with these simulations.

%-----------------------------------------------------------------------------------------------------------------------------
\subsubsection{Number of projectiles}
\label{subsec: numberOfProjectiles}

We arbitrarily considered 10 projectiles in our simulations, but we need to determine whether changing this number has any effect. We therefore re-run the example from Figure \ref{fig: projectileVsPltOnlySims}, but this time varying the number of projectiles. In each case the total mass of projectiles is kept at the original value, but is distributed between different numbers of equal-mass bodies. As in Section \ref{subsec: massOfProjectiles}, we run multiple simulations for each number of projectiles tested; in each case the planet, star and disk parameters are kept constant, whilst the initial projectile orbits are randomised.

\begin{figure}
    % image: projectileAndResonantStirring/images/stirringIncreaseByProjectiles/stirringIncreaseByProjectiles.vsz
    % data: projectileAndResonantStirring/images/stirringIncreaseByProjectiles/data/sim110DifferentProjMassesAndNumbers.ods
    % Run projectileAndResonantStirring/runStirringCheckCodeOnAllSimsFixedPars.py and copy results into ods
    \centering
    \includegraphics[width=8cm]{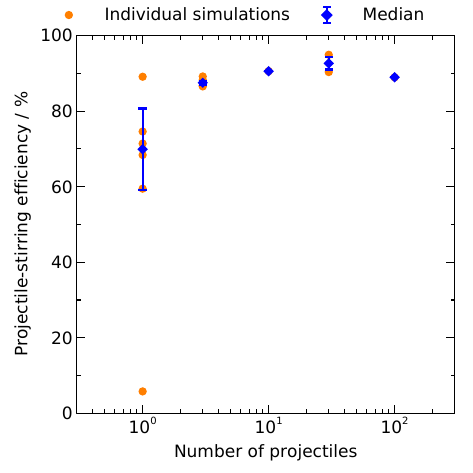}
    \caption{Efficiency of projectile stirring for the example system on Figure \ref{fig: projectileVsPltOnlySims}, as a function of the initial number of projectiles. The total mass of projectiles is always ${0.88\mEarth}$ at the start of each of these simulations, but it is distributed amongst different numbers of equal-mass projectiles. Provided there are enough projectiles, the stirring level is essentially independent of the number of projectiles; the total mass of projectiles is much more important in setting the stirring level (Figure \ref{fig: stirringIncreaseVsProjectileMass}). If only one projectile is present then the results are highly stochastic, because the stirring level is highly senstitive to the chaotic scattering evolution of that projectile.}
    \label{fig: stirringIncreaseVsNumberOfProjectiles}
\end{figure}

Figure \ref{fig: stirringIncreaseVsNumberOfProjectiles} shows the results. The plot shows that, provided the number of projectiles is not too small, the number has minimal effect on the simulation outcome; for this specific setup, the outcome is similar regardless of the number of projectiles, provided it is at least three. The outcome is much more variable if just one projectile is initialised, because in that case the stirring level would be strongly linked to the stochastic evolution of the projectile as it is repeatedly scattered. Comparing Figure \ref{fig: stirringIncreaseVsNumberOfProjectiles} to Figure \ref{fig: stirringIncreaseVsProjectileMass}, it appears that the total mass of projectiles is more important than the number of projectiles in setting the stirring outcome.

%-----------------------------------------------------------------------------------------------------------------------------
\subsection{Analytical prediction of projectile stirring}
\label{subsec: analyticPredOfProjectileStirring}

We now aim to find a general analytical prediction for whether projectile stirring would be important in a debris-disk system.

Since projectile stirring is a secular effect (Section \ref{subsec: isProjStirSecOrScat}), we expect projectile stirring will occur if the projectiles can secularly affect debris before the projectiles are ejected by the planet. We first define the relevant ejection timescale in Section \ref{subsec: particle_scattering}, then the relevant secular timescale in Section \ref{subsec: projStirringTimescaleArgument}, before combining these to produce an analytic criterion in Section \ref{subsec: analyticProjStirringTimescaleCriterion}.

%-----------------------------------------------------------------------------------------------------------------------------
\subsubsection{Scattering of projectiles by the planet}
\label{subsec: particle_scattering}

We first confirm that the diffusion timescale is the best means to quantify how long it takes projectiles to be scattered and ejected by the planet. We have so far argued that particles initialized within \mbox{3 Hill} radii of a planet are likely to be scattered and ultimately ejected, and that this is characterised by the diffusion timescale; here we test both of these assumptions across all of our simulated systems.

\begin{figure}
    \centering
    \includegraphics[width=8cm]{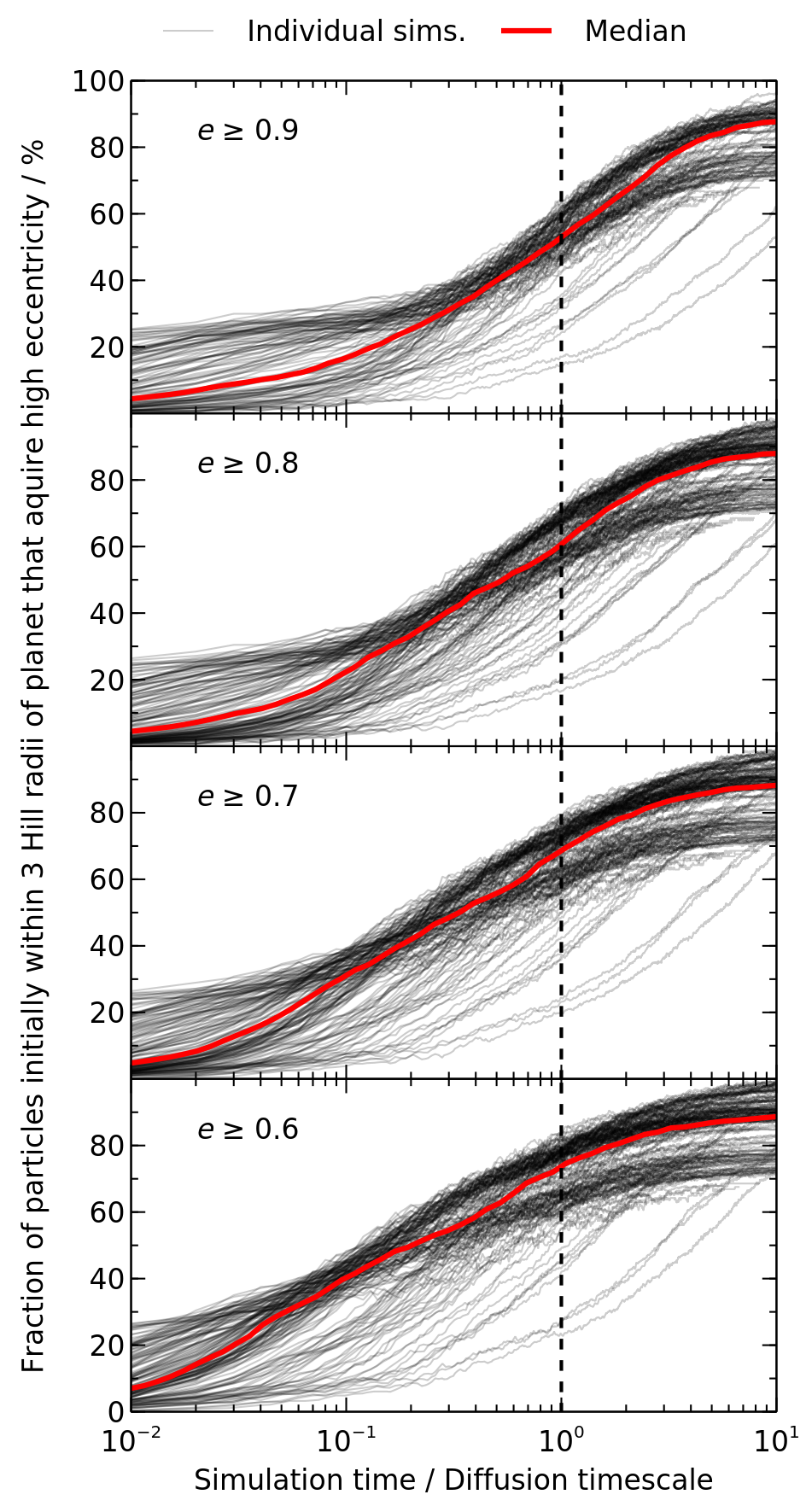}
    \caption{Percentage of massless particles initially within 3 Hill radii of a planet that are ejected or scattered, as a function of time (expressed in diffusion timescales) for all of our planet-only simulations. Panels show the fraction of particles acquiring eccentricities above the given value. In all cases low-eccentricity test particles ${(e \leq 10^{-4})}$ were initially distributed with semimajor axes spanning from the planet's semimajor axis out to at least 5 Hill radii (with an initial surface density profile going as ${r^{-1.5}}$). No massive projectiles were present. Based on the results of these simulations, at 1 diffusion timescale it is expected that between 50-70\% of test particles initially within 3 Hill radii of the planet will already be scattered. The simulations where scattering takes much longer than 1 diffusion timescale are those where ${\mPlt/\mStar \gtrsim 0.01}$; for these high mass ratios, the diffusion-timescale approximations begin to break down \citep{Pearce2023}.}
    \label{fig: fractionDebrisScatteredVsTime}
\end{figure}

To gain general insight into scattering, we consider our planet-only simulations. For each simulation, we examine all of the massless debris particles that were initialised within \mbox{3 Hill} radii of the planet. We then plot the fraction of those particles that are ejected or scattered to high eccentricity, as a function of time. The results for all simulations are shown in Figure \ref{fig: fractionDebrisScatteredVsTime}, where the time in each simulation is expressed in terms of the diffusion timescale (Equation \mbox{\ref{eq: diffusion_time}}, evaluated at ${a=\aPlt}$).

Figure \ref{fig: fractionDebrisScatteredVsTime} shows that the diffusion timescale is a good metric to quantify the scattering timescale. After \mbox{1 diffusion} timescale, roughly ${70\percent}$ of particles initialised on low-eccentricity orbits within \mbox{3 Hill} radii of a planet have been scattered up to eccentricities of at least 0.6, and roughly ${50\percent}$ have reached eccentricities of at least 0.9. Both of these percentages include particles ejected from the system. The scattered fractions are much smaller after just \mbox{0.1 diffusion} timescales, being roughly 40 and ${20\percent}$ respectively, whilst after \mbox{10 diffusion} timescales they are both about ${90\percent}$. Note that the ejection fraction may not necessarily reach ${100\percent}$, because some particles initialised near the planet can occupy stable resonant orbits (e.g. particles in the 1:1 MMR in \mbox{Figure \ref{fig: projectileVsPltOnlySims}}). This exercise demonstrates that the diffusion time well-quantifies the timescale for scattering by one planet over a broad range of system parameters, and so should be a good estimate for projectile lifetimes.

%-----------------------------------------------------------------------------------------------------------------------------
\subsubsection{Defining the projectile secular timescale}
\label{subsec: projStirringTimescaleArgument}

We now define the relevant secular timescale for our prediction of projectile-stirring efficiency. The timescale over which a projectile would secularly perturb a debris particle is given by Equation \ref{eq: secular_time} (replacing the planet parameters with those of the projectile); however, evaluating this timescale is not straightforward, because it strongly depends on the semimajor axes of the debris and projectile. The projectile semimajor axis varies greatly throughout a simulation as it is repeatedly scattered by the planet, and the debris spans a range of semimajor axes. To proceed, we must choose ``typical'' values for these to use in our analytic prediction.

First, we choose to consider debris at the outer edge of the initial disk, i.e. setting ${\aDeb \equiv \aOut}$ in Equation \ref{eq: secular_time} (where $\aOut$ is the semimajor axis of the outermost debris body). This is because we are interested in stirring the entire disk and, since processes at the outer edge are typically slower than at the inner edge, if the outer edge is stirred then the inner edge would probably be stirred too. Second, for the projectile semimajor axis we choose a value between the disk inner and outer edges, because the projectiles start at the inner edge and are then scattered outwards. We arbitrarily choose to evaluate the secular time assuming a projectile semimajor axis of 95\% of the distance from the initial inner edge to the outer edge (i.e. a projectile semimajor axis of ${a^\prime \equiv 0.05\aPlt + 0.95\aOut}$); we will later show that this value lets us accurately predict when projectile stirring will occur. Finally, we must decide what mass to use when calculating the secular timescale; we choose to use the total mass of all projectiles (rather than an individual projectile) when evaluating Equation \ref{eq: secular_time}, because we find that the stirring level depends more on the total-projectile mass than on the number of projectiles (Figures \ref{fig: stirringIncreaseVsProjectileMass} and \ref{fig: stirringIncreaseVsNumberOfProjectiles}). So to summarise, when evaluating the secular timescale (Equation \ref{eq: secular_time}) for our analytic prediction of projectile stirring, we set $\mPlt$ to be the total-projectile mass, ${\aDeb \equiv \aOut}$, and consider a projectile semimajor axis of ${a^\prime \equiv 0.05\aPlt + 0.95\aOut}$.

%-----------------------------------------------------------------------------------------------------------------------------
\subsubsection{Analytical criterion for projectile stirring}
\label{subsec: analyticProjStirringTimescaleCriterion}

\begin{figure}
    % projectileAndResonantStirring/images/stirringIncreaseByProjectiles/stirringIncreaseByProjectiles.vsz
    \centering
    \includegraphics[width=8cm]{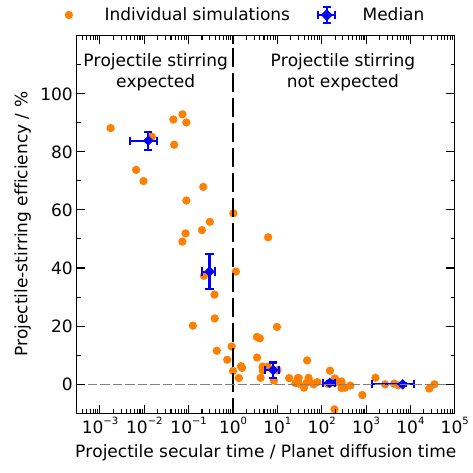}
    \caption{Predicting the efficiency of projectile stirring. Orange circles show the measured projectile-stirring efficiency (Equation \ref{eq: stir_percent_increase}) for each of our N-body simulations with a low-eccentricity planet, as functions of the ratio of the estimated projectile secular time to the planet diffusion time. Simulations where this ratio is less than unity are expected to show projectile stirring, whilst those with ratios greater than unity are not (Section \ref{subsec: projStirringTimescaleArgument}); this agrees with the simulations. Blue points show the binned data.}
    \label{fig: stirringIncreaseByProjectiles}
\end{figure}

Using the results of the previous two sections, we now produce an analytic criterion for when projectile stirring should be important. We argued that projectiles could stir a disk if the secular timescale over which projectiles excite debris, $\tSecProj$, is shorter than the diffusion timescale over which the planet would eject projectiles, $t_{\rm diff,plt}$. In Figure \ref{fig: stirringIncreaseByProjectiles} we plot the projectile-stirring efficiency measured from each of our $N$-body simulations, as a function of the ratio ${\tSecProj/t_{\rm diff,plt}}$ (evaluated for each simulation using the assumptions in Sections \ref{subsec: particle_scattering} and \ref{subsec: projStirringTimescaleArgument}). The figure shows that the theoretical criterion ${\tSecProj < t_{\rm diff,plt}}$ is a good indicator of whether projectile stirring will occur.

We can substitute the expressions for the two timescales in the criterion ${\tSecProj < t_{\rm diff,plt}}$. The result is that projectile stirring is predicted to occur if the total mass of projectiles satisfies

\begin{equation}
    \begin{aligned}
        & \left(\frac{\mProjTot}{{\mEarth}}\right) \gtrsim 120 \left(\frac{a^\prime}{\rm au}\right)^{-1} \left(\frac{\aPlt}{\rm au}\right)^{-3/2} \left(\frac{\aOut}{\rm au}\right)^{5/2}  \\
        & \times \left(\frac{\mPlt}{\mJup}\right)^{2}   \left(\frac{\mStar}{{\mSun}}\right)^{-1} \left[ b_{\rm 3/2}^{(1)} \left(\frac{a^\prime}{\aOut}\right) \right]^{-1},
    \end{aligned}
    \label{eq: theoMinProjMassToStir}
\end{equation}

\noindent where ${a^\prime \equiv 0.05~\aPlt + 0.95~\aOut}$, and the Laplace coefficient $b_{\rm 3/2}^{(1)}(\alpha)$ is given by \mbox{Equation \ref{eq: laplace_coeff}} (a code to evaluate Laplace coefficients is available online\footnote{\href{https://www.tdpearce.uk/public-code/}{tdpearce.uk/public-code}}).

For the simulation in the right of \mbox{Figure \ref{fig: projectileVsPltOnlySims}}, \mbox{Equation \ref{eq: theoMinProjMassToStir}} predicts a total-projectile mass of at least ${0.040\mEarth}$ is needed to stir the disk. This is consistent with the simulation; the total-projectile mass in that simulation is much higher (${0.88\mEarth}$), and indeed the simulated disk is stirred. This theoretical prediction is also shown in Figure \ref{fig: stirringIncreaseVsProjectileMass}, and it agrees with projectile mass at which those simulations transition from projectile stirring being inefficient to efficient.

\begin{figure}
    % projectileAndResonantStirring/images/stirringIncreaseByProjectiles/stirringIncreaseByProjectiles.vsz
    \centering
    \includegraphics[width=8cm]{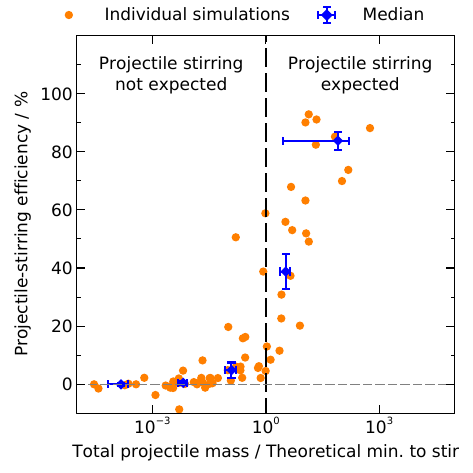}
    \caption{Verification of Equation \ref{eq: theoMinProjMassToStir}, the theoretical minimum total mass of projectiles required to stir a debris disk. The vertical axis shows the projectile-stirring efficiencies from our N-body simulations, and the horizontal axis is the total mass of all projectiles in each simulation, divided by the theoretical minimum to stir that disk (Equation \ref{eq: theoMinProjMassToStir}). Simulations with total-projectile masses above the theoretical minimum tend to have significant projectile stirring, whilst those with smaller total masses tend to exhibit little projectile stirring.}
    \label{fig: stirringIncreaseVsPredMass}
\end{figure}

To verify Equation \ref{eq: theoMinProjMassToStir} across all our simulations, in Figure \ref{fig: stirringIncreaseVsPredMass} we plot our measured projectile-stirring efficiencies, against the ratio of the actual projectile mass to the Equation \ref{eq: theoMinProjMassToStir} prediction. The plot shows that, in simulations where the total-projectile mass is less than the theoretical requirement from Equation \ref{eq: theoMinProjMassToStir}, projectile stirring is absent or minimal; conversely, projectile stirring is significant in simulations where the total-projectile mass exceeds the theoretical requirement. Hence Equation \ref{eq: theoMinProjMassToStir} is a good indicator of the total-projectile mass required to stir a disk, and can be used to predict the outcome of projectile stirring.

Equation \ref{eq: theoMinProjMassToStir} lets us identify the properties of systems where projectile stirring is more likely to occur. Projectile mass is a key factor; larger projectiles are better able to stir debris. Projectile stirring is also more likely if the mass of the scattering planet is small, because smaller planets take longer to eject projectiles, and hence allow them more time to stir. However, if the planet mass is too small, then it may take too long to scatter projectiles in the first place; specifically, if the planet's diffusion time (Equation \ref{eq: diffusion_time}) is much longer than the system age, then it is unlikely that the planet will have yet scattered projectiles onto high-eccentricity orbits. Projectile stirring is also more efficient around high-mass stars, because this increases the ratio of the diffusion and secular timescales. Finally, projectile stirring is most efficient in small, narrow debris disks.

%#############################################################################################################################
\section{Resonant stirring}
\label{sec: resonant_stirring}

In the previous sections, we analysed projectile stirring by comparing the debris-excitation level in N-body simulations that contained a planet, projectiles and debris to equivalent simulations without projectiles. However, we noticed that many simulations without projectiles were stirred to a much greater degree than would be expected from secular-planet-stirring, despite the planet being the only possible driver of stirring. In these simulations, it transpired that stirring was being performed by the planet's mean-motion resonances, which excite material across a much larger region of the disk than expected. In this section we discuss ``resonant stirring'' as a means to increase planet-stirring efficiency. We plan to examine this effect in greater detail in a future publication, including analytic predictions of resonant-stirring efficiency; for the current paper, we demonstrate the effect and discuss its prevalence in our N-body simulations.

In Section \ref{subsec: resStirringExamples} we present example N-body simulations of resonant stirring, and in Section \ref{subsec: identifyingMMRsAsTheStirrers} we demonstrate that mean-motion resonances are the stirring mechanism. In Section \ref{subsec: mmrsAcrossAllSims} we analyse resonant stirring across all of our simulations, and demonstrate that it seems viable for planet masses above ${\sim0.5\mJup}$.

%-----------------------------------------------------------------------------------------------------------------------------
\subsection{Example resonant-stirring simulations}
\label{subsec: resStirringExamples}

\begin{figure*}
    % projectileAndResonantStirring/images/nBodyElements/nBodyElements.vsz
    % This is sim 67
    \includegraphics[width=17cm]{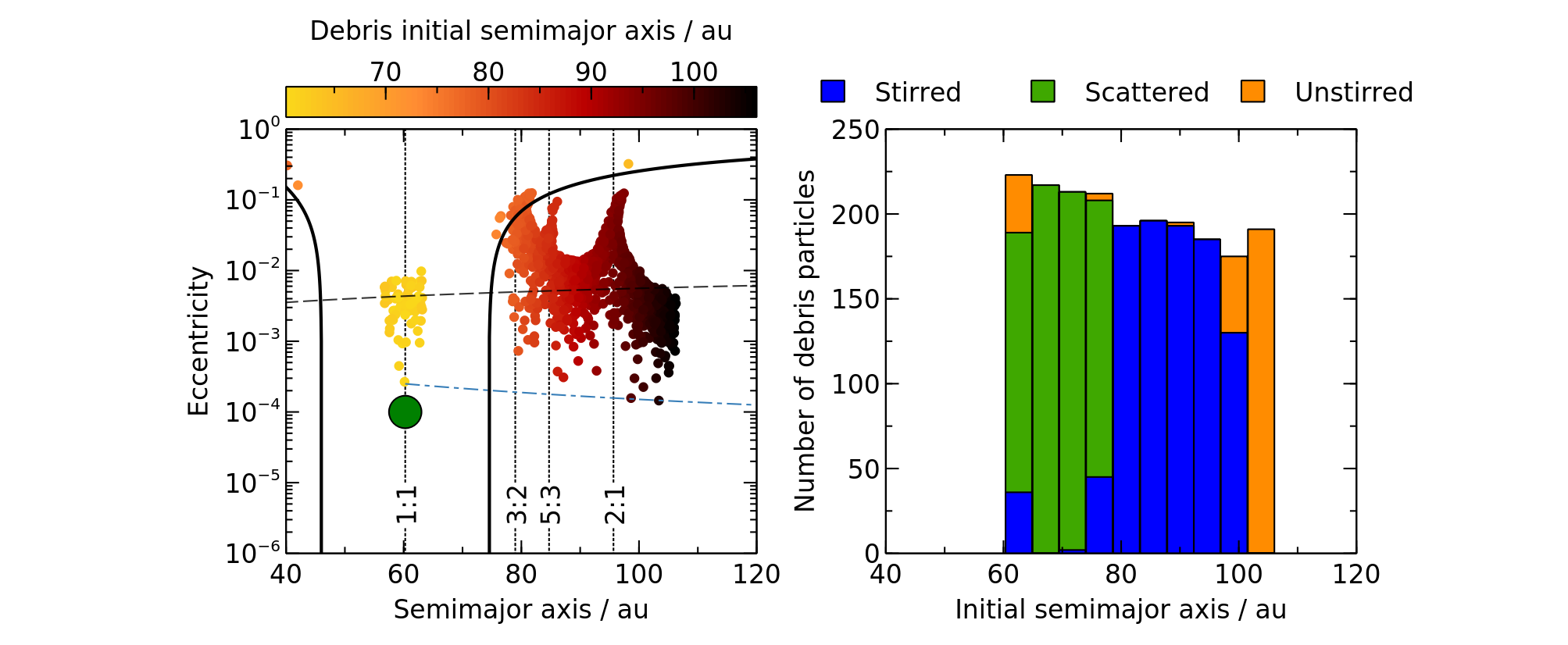}
    \caption{Resonant stirring by a planet. Since mean-motion resonances are broad in semimajor-axis space, they can stir debris not only at nominal MMR locations, but between them too. This can stir a very wide region of the disk. The plot shows a simulation with a ${1.3\mSun}$ star, a ${1.9\mJup}$ planet with eccentricity $10^{-4}$, a massless debris disk and no massive projectiles; the simulation is shown after ${20\myr}$ (\mbox{10 diffusion} timescales). The entire region interior to the 2:1 MMR has been stirred by broad MMRs alone. All lines and symbols were defined on Figure \ref{fig: projectileVsPltOnlySims}.}
    \label{fig: resStirringHighMassPlanet}
\end{figure*}

\begin{figure*}
    % projectileAndResonantStirring/images/nBodyElements/nBodyElements.vsz
    % This is sim 67_0p1MPlt
    \centering
    \includegraphics[width=17cm]{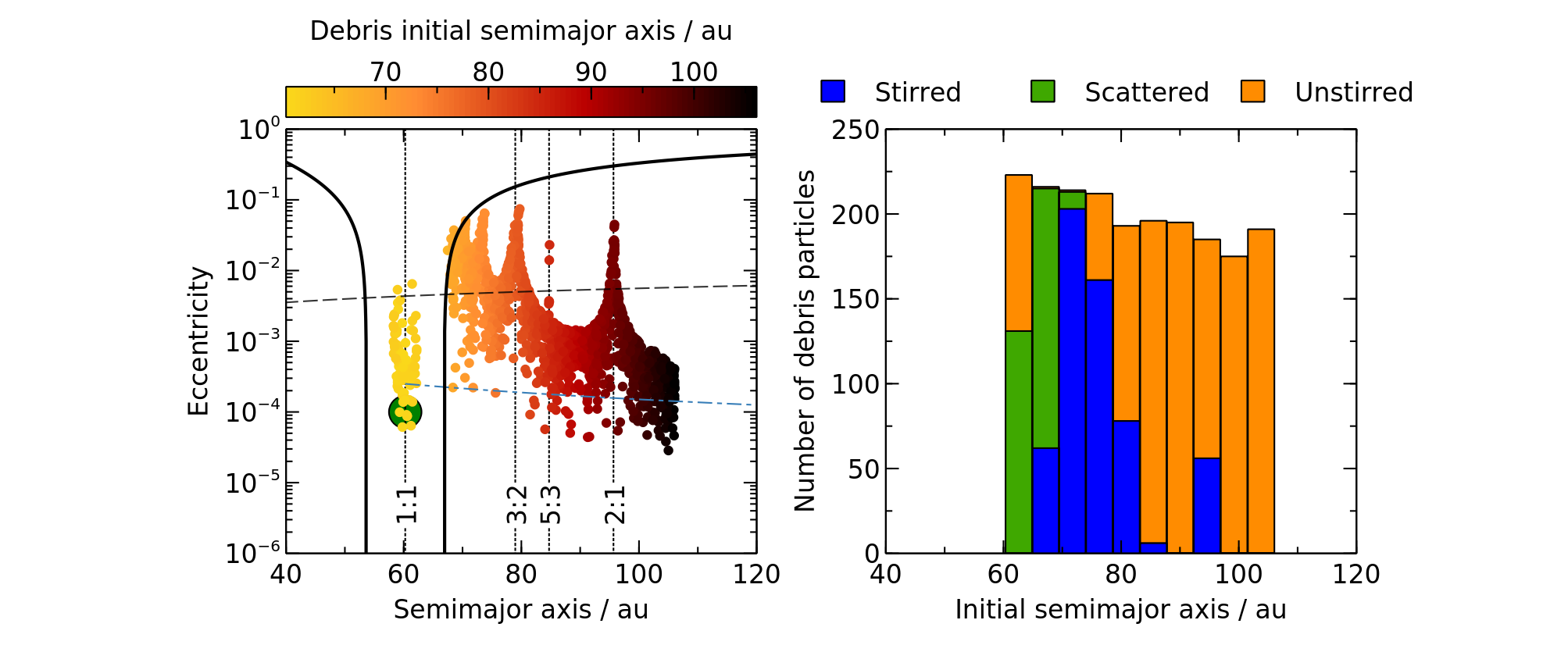}
    \caption{A case where resonant stirring is unable to stir the majority of a debris disk. The simulation setup is identical to that in Figure \ref{fig: resStirringHighMassPlanet}, except that this planet is one order of magnitude less massive (${0.19\mJup}$). The simulation is shown after ${2\gyr}$ (\mbox{10 diffusion} timescales).}
    \label{fig: resStirringLowMassPlanet}
\end{figure*}

Figure \ref{fig: resStirringHighMassPlanet} shows an example N-body simulation where resonant stirring occurs. The system comprises a ${1.3\mSun}$ star, a ${1.9\mJup}$ planet at ${60\au}$, and a debris disk initially extending from the planet out to ${110\au}$. By the end of the simulation (\mbox{10 diffusion} timescales), the planet has ejected non-resonant debris originating within \mbox{3 Hill} radii of its orbit, but debris beyond this is stirred across almost the entire disk, despite no projectiles being present. Specifically, the majority of debris particles are excited to eccentricities above $10^{-2}$ to $10^{-1}$, compared to their initial eccentricities of less than $10^{-4}$. The stirring level is linked to the planet's mass; Figure \ref{fig: resStirringLowMassPlanet} shows a simulation with an identical setup but with the planet mass reduced by a factor of 10 (to ${0.19\mJup}$), and the stirring level is greatly reduced.

%-----------------------------------------------------------------------------------------------------------------------------
\subsection{Identifying mean-motion resonances as the stirring mechanism}
\label{subsec: identifyingMMRsAsTheStirrers}

We now demonstrate that MMRs are the stirring mechanism in the planet-only simulation in Figure \ref{fig: resStirringHighMassPlanet}. In such simulations comprising a star, a planet and massless debris, there are only three interactions that could excite debris: scattering interactions, secular interactions, and mean-motion resonances. We assess each of these here.

Planet-debris scattering can be quickly discounted as the stirring mechanism. This is because scattering exchanges both energy and angular momentum, so it would change debris semimajor axes as well as eccentricities; however, the semimajor axes of surviving debris in the Figure \ref{fig: resStirringHighMassPlanet} simulation remain essentially unchanged (as denoted by particle colours). In addition, the outer edge of the debris disk is at ${110\au}$, which is \mbox{9.7 Hill} radii beyond the planet; this is far outside the expected scattering region.

The second potential stirring mechanism is a secular interaction between the planet and debris; this is the ``traditional'' planet-stirring mechanism considered in the literature \citep{mustill2009}. Unlike scattering, secular interactions conserve energy, so would be consistent with the debris semimajor axes remaining unchanged in the simulation. However, secular stirring can also be discounted, because the planet is not eccentric enough. For debris with initially low eccentricity, an eccentric planet would cause the debris eccentricity to oscillate between roughly its initial value and twice the forcing eccentricity $\eForced$, where 

\begin{equation}
    2 \eForced(\aDeb) = 2\times \frac{5}{4}\frac{\aPlt}{\aDeb}\ePlt.
    \label{eq: forced_ecc}
\end{equation}

\noindent The forcing eccentricity, and hence the maximum eccentricity attained by secular debris, is set by the planet eccentricity and the planet-debris-semimajor-axis ratio. For the simulation in Figure \ref{fig: resStirringHighMassPlanet}, the the maximum debris eccentricity expected at the disk outer edge from secular interactions is therefore

\begin{equation}
    \centering
    2 \eForced(110\au) = 2 \times \frac{5}{4} \frac{60\au}{110\au} \times 10^{-4} = 1.4\times10^{-4},
    \label{eq: forced_ecc_calculated}
\end{equation}

\noindent which is far smaller than the eccentricities of ${4\times10^{-3}}$ at the outer edge of the simulated disk. Even nearer to the planet, where the forcing eccentricity would be higher, debris is still much more eccentric than would be expected from secular interactions; the dot-dash line on Figure \ref{fig: resStirringHighMassPlanet} shows ${2\eForced}$ across the disk, which is far smaller than the actual debris-excitation level. Hence secular interactions can also be discounted as the stirring mechanism.

Mean-motion resonances are therefore the only possible mechanism that could have stirred the disk. MMR interactions cause debris to oscillate in eccentricity and semimajor axis over a narrow range, so are again consistent with debris semimajor axes being essentially unchanged in the simulation. From Figure \ref{fig: resStirringHighMassPlanet} it is clear that MMRs are operating; debris has well-defined eccentricity peaks just outside the nominal 3:2, 5:3 and 2:1 MMRs, as expected from resonant interactions. Importantly, these peaks are not infinitely narrow in semimajor-axis space; they have width, and the maximum debris eccentricity smoothly decreases at semimajor axes on either side of the nominal MMR locations. This width is a fundamental property of MMRs, and the shapes of the debris profiles in eccentricity-semimajor-axis space around e.g. the nominal 2:1 MMR on Figure \ref{fig: resStirringHighMassPlanet} well match those expected from MMRs \citep{murray1999}. We therefore conclude that mean-motion resonances are responsible for stirring the disk in this case, and importantly, that their non-zero widths mean that MMRs can stir debris over a much broader region than just their nominal locations.

%-----------------------------------------------------------------------------------------------------------------------------
\subsection{Resonant stirring across our simulations}
\label{subsec: mmrsAcrossAllSims}

Resonant stirring occurs in at least one third of our planet-only simulations (i.e. those without projectiles). By this, we mean that the debris exhibits the characteristic eccentricity-semimajor-axis profile seen in Figure \ref{fig: resStirringHighMassPlanet}, and a substantial region of the disk is stirred (not only at nominal MMR locations, as in Figure \ref{fig: resStirringLowMassPlanet}, but between them too). In this section we analyse resonant stirring across our planet-only simulations; this effect is probably also present in our projectile simulations, but we omit those from this analysis to avoid confusing the effects of resonant and projectile stirring. Since MMR theory is more complicated than the relatively simple analyses we performed for projectile stirring (Section \ref{sec: projectileStirring}), we intend to perform a detailed resonant-stirring analysis in a dedicated future paper; in this section, we instead aim to identify the key parameters that set the importance of resonant stirring.

For massless, resonant debris, the main parameters predicted to set the debris eccentricity and MMR width are the planet-to-star-mass ratio $\mPltOverMStar$, and the terms associated with the specific MMR (i.e. $p$ and $q$ in the MMR notation ${p+q:p}$; \citealt{murray1999, Petrovich2013, Pearce2023}). Typically, higher planet-to-star-mass ratios and lower-order resonances result in higher debris eccentricities and broader MMRs. We therefore expect eccentricity excitation to be highest for high-mass planets, and for disks near the strong 3:2 and/or 2:1 MMRs. This is consistent with our simulations; Figures \ref{fig: resStirringHighMassPlanet} and \ref{fig: resStirringLowMassPlanet} show that debris is most excited around these two MMRs, and the degree of excitation is higher for the higher-mass planet.

However, $p$, $q$ and $\mPltOverMStar$ alone are not sufficient to determine whether resonant stirring would occur. The reason is that the debris eccentricity required for stirring also depends on other system parameters; specifically, Equations \ref{eq: unstirred_ecc} to \ref{eq: final_unstirred_ecc} show that the minimum eccentricity for stirring depends on the debris material, disk location and star mass. Hence the resonant-stirring efficiency may differ in systems with identical $\mPltOverMStar$ and relative debris locations, but different star masses and/or planet locations. We therefore hypothesise that the main parameters setting the efficiency of resonant stirring are the star mass, planet mass, locations of the disk edges, and debris composition.

We now test resonant-stirring efficiency against these parameters, to establish which are most important. We first define a rough proxy for resonant-stirring efficiency in our planet-only simulations. Since we define resonant stirring to have occurred if debris is stirred both at and between nominal MMR locations, and since the 3:2 and 2:1 MMRs are typically the strongest, we choose to take the stirring level at the midpoint between the nominal 3:2 and 2:1 locations as a proxy for the degree of resonant stirring in a planet-only simulation. For this analysis we exclude any simulations with massive projectiles, or where the midpoint between the 3:2 and 2:1 MMRs lies beyond the disk outer edge, or where the midpoint lies within \mbox{5 Hill} radii of the planet (to exclude cases where debris is excited by scattering rather than MMRs). For each of the included simulations, we run our numerical stirring analysis (\mbox{Section \ref{subsec: numericalStirringAnalysis}}), then bin each debris particle into one of 20 equal-width bins depending on its initial semimajor axis. Following this process, if there are more than 10 debris particles in the bin coinciding with the midpoint between the 3:2 and 2:1 MMRs, then we use the fraction of stirred particles in this bin as our proxy for the degree of projectile stirring in the simulation.

\begin{figure}
    % projectileAndResonantStirring/images/stirringIncreaseByProjectiles/stirringIncreaseByProjectiles.vsz
    \centering
    \includegraphics[width=8cm]{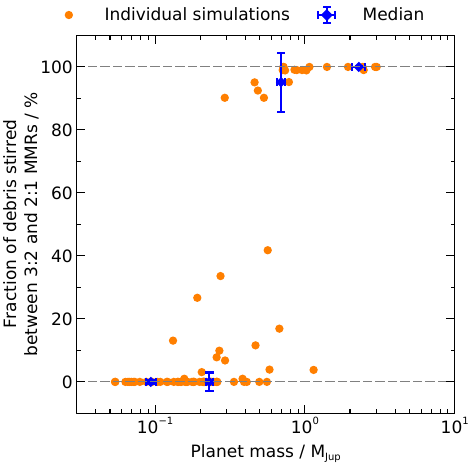}
    \caption{Efficiency of resonant stirring versus planet mass. The vertical axis shows the stirring level at the midpoint between the 3:2 and 2:1 MMRs, which we use as a proxy for resonant stirring, measured from a large fraction of our planet-only N-body simulations (i.e. those without projectiles). Planets with masses above ${\sim0.5\mJup}$ can stir debris through MMRs alone, and this critical planet mass seems to be relatively independent of other system parameters (star mass, disk location etc.). Resonant stirring has a less-clear dependence on those other parameters than on planet mass (Figure \ref{fig: resStirringAppGrid}).}
    \label{fig: resonantStirringVsPlanetMass}
\end{figure}

Figure \ref{fig: resonantStirringVsPlanetMass} shows the results of this analysis as a function of planet mass. Of the parameters listed above, planet mass appears to be the most important in setting the efficiency of resonant stirring; for our tested parameter space (Table \ref{tab: random_sim_setup}), it appears that resonant stirring is significant for planet masses above ${\sim0.5\mJup}$. This critical planet mass appears to be relatively independent of star mass and disk-edge locations. Other parameters do not seem to affect resonant stirring as clearly; we show the resonant-stirring proxy as a function of star mass, planet-to-star-mass ratio, and disk-edge locations in \mbox{Figure \ref{fig: resStirringAppGrid}}, but these individual parameters do not have as tight relations to resonant-stirring efficiency as planet mass does. 

In summary, broad mean-motion resonances from an internal planet can efficiently stir debris disks, and we show that this effect can become significant for planet masses above ${\sim0.5\mJup}$. We identify this mechanism as a potential means to significantly increase planet-stirring efficiency, especially for low-eccentricity planets. However, we do not conduct a detailed numerical or theoretical exploration of resonant stirring here; that analysis is planned for a future paper.

%#############################################################################################################################
\section{Discussion}
\label{sec: discussion}

We demonstrated that the efficiency of planet-stirring can be increased through two additional mechanisms: projectile stirring and resonant stirring. In this section we discuss how projectile and resonant stirring fit in with the other stirring mechanisms (Section \ref{subsec: discussionImplicationsOfProjAndResStirring}), the viability of projectile stirring (Section \ref{subsec: discussionViabilityOfProjAndResStirring}), the limitations of our analyses (Section \ref{subsec: discussionLimitationsOfAnalyses}), and what future studies could focus on (Section \ref{subsec: discussionFutureDirections}).

%-----------------------------------------------------------------------------------------------------------------------------
\subsection{Implications of projectile and resonant stirring}
\label{subsec: discussionImplicationsOfProjAndResStirring}

The nature of debris stirring is unknown. Several mechanisms have been proposed, but these have problems; secular-planet-stirring requires the planet to have high eccentricity (\citealt{mustill2009}), which is not necessarily consistent with known Solar-System and extrasolar planets. It may also run into difficulty when disk mass is considered, since disk self-gravity may resist secular-planet-stirring (L\"{o}hne et al., in prep.). The most widely considered alternative, self-stirring due to debris-disk self gravity, also has significant problems; it cannot be occurring in some debris disks unless their size distributions are very different to conventional debris theory (\citealt{krivov2021, najita2022, pearceispy2022}).

We showed that planet-stirring could be significantly more efficient than previously thought, due to the additional effects of projectiles and mean-motion resonances. In particular, these effects would occur even for non-eccentric planets. They add to the arsenal of mechanisms by which planets could stir debris; as well as secular stirring by eccentric planets \citep{mustill2009}, these include stirring by MMR sweeping as planets migrate \citep{friebe2022}, and interplays between planet-debris interactions and disk self gravity (`mixed stirring'; \citealt{munoz2015dpdiskexcitation, munozmixedstirring2023}). These mechanisms open the potential for planet-stirring to be more plausible than previously thought, with the possibility that planets are the main debris-stirring mechanism in nature.

Planet-stirring could potentially be tested soon; the \textit{James Webb Space Telescope} (\textit{JWST}) is searching for planets near the inner edges of debris disks, and detections (or non detections) would help assess the viability and prevalence of planet-stirring. Our paper constrains the planet masses required for projectile and/or resonant stirring to occur, which could be used to interpret \textit{JWST} planet searches. If no sufficiently massive planets were found, then this could point towards the prevalence of less-considered stirring mechanisms; examples include pre-stirring and flyby stirring.

Finally, there is a potential implication of projectile stirring for the Solar System. Applying Equation \ref{eq: theoMinProjMassToStir} to the Kuiper Belt and Neptune, we see that projectile stirring of the Belt is possible if projectiles had a total mass of at least ${4 \times 10^{-4}\mEarth}$. This is just ${10 \percent}$ of the mass of the dwarf planet Eris which, as noted in Section \ref{sec: introduction}, has a highly eccentric, Kuiper-Belt crossing orbit, consistent with historical scattering by Neptune. This raises the possibility that the Kuiper Belt may have been dynamically excited by projectile stirring. Neptune's diffusion timescale is ${0.6\gyr}$ (Equation \ref{eq: diffusion_time}), which is also an upper limit on the secular timescale of projectiles with a total mass of at least ${4 \times 10^{-4}\mEarth}$. Hence Neptune has had enough time to scatter projectiles, and those projectiles have had enough time to stir the Kuiper Belt, within the age of the Solar System. Similar ideas have been suggested before; for example, secular perturbations by an Earth-mass rogue planet could have excited Kuiper-Belt Objects (KBOs) and generated the detached population \citep{Gladman2006, Huang2022}. Our results suggest that Kuiper-Belt stirring could have been performed by smaller dwarf planets, like Eris, which were scattered by Neptune. However, we are cautious with this conclusion, because we have not performed specific Solar-System modelling; in particular, we do not simulate Neptune's migration, nor include the Kuiper-Belt mass, nor attempt to reproduce the specific populations of resonant and detached KBOs. Nonetheless, the existence of high-eccentricity Eris, combined with the dynamical excitation of many KBOs, could imply that some form of projectile stirring has occurred in our own Solar System.

%-----------------------------------------------------------------------------------------------------------------------------
\subsection{Viability of projectile and resonant stirring}
\label{subsec: discussionViabilityOfProjAndResStirring}

There are several requirements of a planetary system in order for projectile stirring to occur. First, a planet must be present, with a supply of nearby, massive bodies to scatter. This appears plausible, based on our knowledge of the Solar System and extrasolar systems; Neptune scattered a large quantity of massive debris from the inner edge of the Kuiper Belt \citep{Gomes2008}, and many extrasolar debris disks have sharp inner edges and/or extended radial profiles indicative of planetary sculpting and/or scattered populations \citep{geiler2019, Faramaz2021, matra2019, ImazBlanco2023, Pearce2023}. This implies that, in at least some cases, planets are (or were) within a few Hill radii of debris populations, which they then scattered. For our simulations, we imposed a scattered population by initializing the planet at the disk inner edge, but it is not a requirement of projectile stirring that planets formed in such locations; for example, a planet forming away from a debris disk and then migrating towards it could also result in projectile stirring. Since we have evidence of such planetary migration in both the Solar System \citep{Gomes2005, Morbidelli2005, Tsiganis2005} and possibly extrasolar debris-disk systems \citep{friebe2022, Booth2023}, post-formation migration is a plausible means to initiate projectile stirring. Hence the first requirement for projectile stirring, a supply of massive bodies close to a planet, appears plausible.

The second requirement is that the total mass of projectiles must be sufficiently high. Equation \ref{eq: theoMinProjMassToStir} shows that, for a debris disk centred on ${100\au}$ with width ${50\au}$, the required combined mass of projectiles is ${\sim 0.16 (\mPlt/\mJup)^2 (\mStar/\mSun)^{-1}}$. So if, for example, a Jupiter-mass planet were present at the disk inner edge, and the star were A-type (${\sim 2 \mSun}$), then the required total mass of projectiles would be ${\sim30\mEarth}$. This mass is plausible; the early Kuiper Belt is thought to have had a mass of ${\sim10\mEarth}$ \citep{Nesvorny2018, Gladman2021}, and extrasolar debris disks could have masses up to $100-1000~{\mEarth}$ \citep{krivov2021}. Hence sufficiently massive projectile reservoirs should be available. Taken together, the two main requirements for projectile stirring seem likely to be satisfied in at least some systems; based on this, we argue that projectile stirring is a viable means to stir debris disks.

For resonant stirring, the requirements should be even easier to satisfy. The minimum-required planetary mass is just ${\sim0.5\mJup}$, and such planets are known to exist in our Solar System and beyond. Furthermore, the planet need not even be located close to the debris disk; in the example on Figure \ref{fig: resStirringHighMassPlanet}, debris is stirred all the way out to the nominal location of the 2:1 MMR. In such a system, debris with semimajor axis $a$ would be stirred if ${a \leq 1.6\aPlt}$, so even well-separated planets could stir debris disks. Whilst Jupiter and Saturn may hence be too distant to stir the Kuiper Belt, there are known extrasolar systems with massive planets sufficiently close to debris-disk inner edges (e.g. ${\HR8799}$; \citealt{Marois2008, Faramaz2021}). Hence the requirement of a sufficiently massive planet close enough to a debris disk seems plausible from currently known planets, and there may also be additional planets even closer to disks that are yet to be discovered. In addition, were a planet to migrate, then MMR sweeping would increase both the trapping efficiency and eccentricity of debris \citep{friebe2022, Booth2023}, so migration would further enhance resonant stirring. In summary, both resonant and projectile stirring appear to be viable mechanisms to stir debris disks, with the required conditions thought to be satisfied in at least some systems.

%-----------------------------------------------------------------------------------------------------------------------------
\subsection{Limitations of our analyses}
\label{subsec: discussionLimitationsOfAnalyses}

We now discuss the limitations of our numerical and analytic analyses, and their potential impact on our results.

%-----------------------------------------------------------------------------------------------------------------------------
\subsubsection{Debris-disk masses}

We modelled most debris as massless; only a few projectiles were assigned masses in our N-body simulations. This meant we neglected disk self-gravity, particularly in simulations without massive projectiles, or in the outer regions of the disk (away from projectiles). There are two main effects of this omission. First, our disks are unable to undergo self-stirring, so the stirring level may be underestimated in some cases (especially for broad disks in simulations without projectiles). Second, the disks are unable to resist perturbations from the planet and projectiles; L\"{o}hne et al. (in prep.) show that massive debris disks can resist secular-planet-stirring, and it is therefore possible that they would resist projectile and resonant stirring too. Hence the stirring level could be overestimated in some of our simulations. 

However, we justify omitting debris-disk masses for several reasons. First, debris-disk masses are unknown, and the plausible, theoretically justified estimates of disk mass span several orders of magnitude \citep{krivov2021}. Therefore, if we were to include disk mass, it is unclear what mass should actually be assumed. Second, we sought to study the effects of projectile and resonant stirring, and this was done by isolating those mechanisms; including disk mass would have led to self-stirring as well, which would have made the effects of projectile and resonant stirring much harder to disentangle. Third, in projectile simulations, it could be argued that disk mass is actually included; debris-disk masses are thought to be dominated by the largest bodies, so the massive projectiles could be thought of as the largest bodies in the disk. This means that we assume the disk mass is initially concentrated at the disk inner edge (where the projectiles are initialised), but this may be a valid approach; since planetesimal-formation efficiency probably decreases with stellar distance (e.g. \citealt{kenyon2008}), and large bodies may also migrate inwards through planetesimal scattering \citep{friebe2022}, it could be expected that the mass of a broad debris disk is concentrated at its inner edge. For these reasons, we argue that we were justified in omitting disk mass. Doing so significantly sped up our N-body simulations, letting us run many more simulations and explore a much larger region of parameter space.

%-----------------------------------------------------------------------------------------------------------------------------
\subsubsection{Assumption of an initially unstable debris disk}

Our N-body simulations were initialised in an unstable configuration, with debris extending all the way to the planet's orbit. This was deliberate, because we sought to generate a scattered-projectile population. The main alternative for investigating projectile stirring would be to assume a more-distant disk, and insert projectiles on eccentric orbits that cross both the planet and disk. However, the latter approach would be hard to justify, for two reasons. First, the projectile orbits would be artificial and less realistic; the distribution of projectile orbits in semimajor-axis, eccentricity, inclination and orientation space would almost certainly have been biased had we initialised them manually, a bias we avoided by letting the planet naturally generate this scattered population. Second, if we had initialised projectiles on high-eccentricity, disk-crossing orbits, then we would have omitted any system evolution that occurred as the projectiles' eccentricities increased from the low-eccentricity orbits they presumably formed on. By initialising projectiles near the planet, we captured the evolution of the disk and planet as projectile eccentricities grew.

To some degree, our setup is physically justified. As noted in Sections \ref{sec: introduction} and \ref{subsec: discussionViabilityOfProjAndResStirring}, we have considerable evidence that debris scattering occurs in nature. Whilst the specific system architectures and histories that lead to this scattering are debatable, the outcome is the same; somehow, significant quantities of debris have come within a few Hill radii of a planet, and been scattered. We choose to model this in the simplest-possible way, by initialising debris near the planet. In reality, it may be that planets form away from debris disks and then migrate towards them (like Neptune in the Nice Model; \citealt{Gomes2005, Morbidelli2005, Tsiganis2005}), but we chose not to model migration because it requires significant assumptions that would have biased our results. Regardless, if migration were rapid (like Neptune in the Nice Model), then the planet would quickly encounter debris, and the result would roughly resemble our simulation setup. Alternatively, debris could form in the disk and diffuse through self-scattering towards the planet, again leading to scattering. In this case, the projectile-stirring outcome would be similar to our simulations; once massive debris encountered the planet, it would be scattered onto eccentric orbits and projectile stirring would begin.

By initialising particles close to the planet, we created a scattered population of massless debris. Before being ejected, this population acquired high-eccentricity orbits that crossed the unperturbed outer regions of the disk. However, this was accounted for in our stirring analyses, so as not to overestimate the stirring level in the outer regions. As explained in Section \ref{subsec: numericalStirringAnalysis} and Appendix \ref{sec: general_stirring_analysis_appendix}, our numerical stirring analysis only compared pairs of debris particles with similar semimajor axes, and whose semimajor axes remained close to their initial values. This ensured that high-eccentricity, massless debris scattered from the inner edge did not contaminate the measured stirring level further out.

%-----------------------------------------------------------------------------------------------------------------------------
\subsubsection{Omission of inclination in the numerical stirring analyses}

With our numerical stirring analysis, we aim to properly account for different debris eccentricities and orientations when assessing stirring levels. This produces a more-accurate picture than the simple, minimum-eccentricity value from Equations \ref{eq: unstirred_ecc}-\ref{eq: final_unstirred_ecc}, and works by assessing collision speeds at orbit-intersection points. However, we implemented the numerical analysis in just two dimensions; mutual inclinations are not considered when calculating whether debris orbits intersect. We omitted inclinations because we would otherwise need additional, non-physical variables in the code; for example, to account for finite numbers of debris particles in our N-body simulations, we would have had to include numerical impact parameters around each orbit when calculating three-dimensional intersection points. The measured stirring levels could be very sensitive to such artificial parameters, so we chose to omit them by not considering mutual inclinations.

This decision should be valid, because the disks are initialised to be very thin, and the non-scattered regions (where stirring is measured) stay relatively thin throughout the simulations. For example, debris at the outer edge of the disk on the right panels of \mbox{Figure \ref{fig: projectileVsPltOnlySims}} gets excited to a maximum inclination of just ${0.6^\circ}$. For discs comprising a fixed number of bodies, an increased inclination spread would reduce the collision probability, because that probability is proportional to the geometric thickness of the disk. This effect could lead to our stirring percentages being slightly overestimated. However, a counterpoint is that mutual inclinations would also increase collision speeds, meaning that our stirring levels could alternatively be slightly underestimated. Either way, for the thin discs we consider, the omission of inclination is unlikely to strongly influence our stirring analyses. Also, since the code only considers pairs of debris bodies with small mutual inclinations, our stirring results are not affected by highly inclined debris that do not intersect the disk.

%-----------------------------------------------------------------------------------------------------------------------------
\subsubsection{Assumed projectile semimajor axis in secular calculations}

In Sections \ref{subsec: projStirringTimescaleArgument} and \ref{subsec: analyticProjStirringTimescaleCriterion} we predicted the outcome of projectile stirring by considering the secular timescale for projectiles to stir a disk. This analytic calculation required a value for the projectile semimajor axis. However, a projectile's semimajor axis constantly varies throughout an N-body simulation, as it is repeatedly scattered by the planet. We therefore had to assume an ``effective'' semimajor axis in our calculations, which accounted for all projectiles at all times. Our chosen value of ${0.05\aPlt + 0.95\aOut}$ is justified because the resulting equations well reproduce the simulations outcomes (\mbox{Figures \ref{fig: stirringIncreaseByProjectiles}} and \ref{fig: stirringIncreaseVsPredMass}), but assuming a single value across all of our simulations could mask dynamical subtleties. However, since our chosen value appears to result in accurate predictions across our broad parameter space, we do not believe this to be a significant issue.

%-----------------------------------------------------------------------------------------------------------------------------
\subsection{Future directions}  
\label{subsec: discussionFutureDirections}

We conducted a preliminary analysis of projectile and resonant stirring. In this section we outline additional analyses that should be performed in the future, to better understand these mechanisms, their prevalence in nature, and their effect on planet-stirring efficiency.

First, whilst we produced an analytic prediction for projectile-stirring efficiency (Equation \ref{eq: theoMinProjMassToStir}), we did not produce an equivalent for resonant stirring. This is because the required MMR theory is more complicated, and such an analysis lies beyond the scope of this project. We plan to produce a detailed theoretical analysis of resonant stirring in a future paper; specifically, we intend to take analytic prescriptions for the shapes of external MMRs in eccentricity-semimajor-axis space, and combine these with theoretical stirring requirements (e.g. \mbox{Equation \ref{eq: unstirred_ecc}}) to derive analytic predictions of resonant-stirring efficiency. These predictions would then be verified against a broad suit of N-body simulations, and could be applied to observed debris-disk systems.

Second, future works should analyse the interplay between projectile, resonant, and self-stirring. This would involve similar analyses to those conducted here, only with massive debris disks. This would let us assess whether disk self gravity increases or reduces stirring efficiency. Similar studies have been conducted by \cite{munoz2015dpdiskexcitation, munozmixedstirring2023}; comparing results such as these to dynamical predictions for projectile and resonant stirring would help us understand potential connections between different stirring mechanisms.

Future investigations could also include planet migration, which has the potential to both increase and decrease the effectiveness of projectile and resonant stirring. For example, migration could enhance projectile stirring by causing the planet to encounter and scatter more projectiles as it moves through a disk, but conversely it could also hinder stirring because potential projectiles could get trapped in sweeping mean-motion resonances and hence become protected from scattering (\citealt{friebe2022}). Since planet migration is a plausible means to bring planets into contact with projectile reservoirs, the effect of migration on these stirring mechanisms should be investigated.

Finally, projectile and resonant stirring should be applied to real debris disks, to assess their viability in such systems, and to make specific predictions for unseen stirring planets. This could be especially useful for debris disks where the current, traditional stirring models do not fit well; in such cases, the new mechanisms of projectile and resonant stirring could offer insights into how debris disks are stirred.

%#############################################################################################################################
\section{Conclusions}
\label{sec: conclusion}

In this paper, we examined two new mechanisms that could increase the effect of planet-stirring.  The first mechanism is projectile stirring, where massive debris bodies that have been scattered onto eccentric orbits by a planet can stir the remaining debris disk. The second is resonant stirring, where a planet (without any massive projectiles) stirs a debris disk through broad mean-motion resonances alone.

To determine the viability of these stirring mechanisms, we ran N-body simulations of debris-disk systems across a broad parameter space. We then analysed these simulations using a bespoke {\sc python} code, which numerically quantifies the degree of stirring in debris-disk systems simulated with {\sc rebound}. We have made this stirring-analysis code publicly available.

We demonstrated that both projectile and resonant stirring have the potential to significantly increase planet-stirring efficiency. In particular, neither requires the planet's orbit to be eccentric, in contrast to the traditional, secular model of planet-stirring.

For projectile stirring, we identified an analytic condition (Equation \ref{eq: theoMinProjMassToStir}) to predict when this type of stirring is likely to occur.  For resonant stirring, we demonstrated that a planet's broad mean-motion resonances can substantially excite debris eccentricity across a wide region of a disk, rather than only near the nominal locations of MMRs. We showed that this stirring mechanism is viable for planet masses above ${\sim0.5\mJup}$.

We conclude that both massive projectiles scattered by a planet, and broad mean-motion resonances, can significantly increase debris eccentricity and hence greatly increase planet-stirring efficiency.  Previous models of stirring, while applicable in some cases, had difficulty describing how some observed disks came to be stirred; the two new mechanisms we investigate could help explain how stirring is achieved in debris disks.

%%%%%%%%%%%%%%%%%%%%%%%%%%%%%%%%%%%%%%%%%%%%%%%%%%
\section*{Acknowledgements}

 We would like to thank Marco Mu{\~n}oz-Guti{\'e}rrez for his thorough review, which substantially improved the quality of this paper. We also thank him, Jonathan Marshall, and Antonio Peimbert for sharing their manuscript before its publication, and for useful discussions regarding alternate forms of stirring.  TDP and AVK were supported by Deutsche Forschungsgemeinschaft (DFG) grants \mbox{Kr 2164/14-2} and \mbox{Kr 2164/15-2}. TDP is also supported by a Warwick Prize Fellowship, made possible by a generous philanthropic donation.

%%%%%%%%%%%%%%%%%%%%%%%%%%%%%%%%%%%%%%%%%%%%%%%%%%
\section*{Data availability}

The data underlying this article will be shared upon reasonable request to the corresponding author.



\bibliographystyle{mnras}
\bibliography{references} % if your bibtex file is called example.bib

\begin{thebibliography}{}
\makeatletter
\relax
\def\mn@urlcharsother{\let\do\@makeother \do\$\do\&\do\#\do\^\do\_\do\%\do\~}
\def\mn@doi{\begingroup\mn@urlcharsother \@ifnextchar [ {\mn@doi@} {\mn@doi@[]}}
\def\mn@doi@[#1]#2{\def\@tempa{#1}\ifx\@tempa\@empty \href {http://dx.doi.org/#2} {doi:#2}\else \href {http://dx.doi.org/#2} {#1}\fi \endgroup}
\def\mn@eprint#1#2{\mn@eprint@#1:#2::\@nil}
\def\mn@eprint@arXiv#1{\href {http://arxiv.org/abs/#1} {{\tt arXiv:#1}}}
\def\mn@eprint@dblp#1{\href {http://dblp.uni-trier.de/rec/bibtex/#1.xml} {dblp:#1}}
\def\mn@eprint@#1:#2:#3:#4\@nil{\def\@tempa {#1}\def\@tempb {#2}\def\@tempc {#3}\ifx \@tempc \@empty \let \@tempc \@tempb \let \@tempb \@tempa \fi \ifx \@tempb \@empty \def\@tempb {arXiv}\fi \@ifundefined {mn@eprint@\@tempb}{\@tempb:\@tempc}{\expandafter \expandafter \csname mn@eprint@\@tempb\endcsname \expandafter{\@tempc}}}

\bibitem[\protect\citeauthoryear{{Aumann} et~al.,}{{Aumann} et~al.}{1984}]{aumann1984thermal}
{Aumann} H.~H.,  et~al., 1984, \mn@doi [\apjl] {10.1086/184214}, \href {https://ui.adsabs.harvard.edu/abs/1984ApJ...278L..23A} {278, L23}

\bibitem[\protect\citeauthoryear{{Backman} \& {Paresce}}{{Backman} \& {Paresce}}{1993}]{backman1993}
{Backman} D.~E.,  {Paresce} F.,  1993, in {Levy} E.~H.,  {Lunine} J.~I.,  eds, Protostars and Planets III. p.~1253

\bibitem[\protect\citeauthoryear{{Beust}}{{Beust}}{2016}]{Beust2016}
{Beust} H.,  2016, \mn@doi [\aap] {10.1051/0004-6361/201628638}, \href {https://ui.adsabs.harvard.edu/abs/2016A&A...590L...2B} {590, L2}

\bibitem[\protect\citeauthoryear{{Beust} et~al.,}{{Beust} et~al.}{2014}]{beust2014}
{Beust} H.,  et~al., 2014, \mn@doi [\aap] {10.1051/0004-6361/201322229}, \href {https://ui.adsabs.harvard.edu/abs/2014A&A...561A..43B} {561, A43}

\bibitem[\protect\citeauthoryear{{Booth} \& {Clarke}}{{Booth} \& {Clarke}}{2016}]{boothclarke2016}
{Booth} R.~A.,  {Clarke} C.~J.,  2016, \mn@doi [\mnras] {10.1093/mnras/stw488}, \href {https://ui.adsabs.harvard.edu/abs/2016MNRAS.458.2676B} {458, 2676}

\bibitem[\protect\citeauthoryear{{Booth} et~al.,}{{Booth} et~al.}{2023}]{Booth2023}
{Booth} M.,  et~al., 2023, \mn@doi [\mnras] {10.1093/mnras/stad938}, \href {https://ui.adsabs.harvard.edu/abs/2023MNRAS.521.6180B} {521, 6180}

\bibitem[\protect\citeauthoryear{{Dermott} \& {Murray}}{{Dermott} \& {Murray}}{1983}]{Dermott1983}
{Dermott} S.~F.,  {Murray} C.~D.,  1983, \mn@doi [\nat] {10.1038/301201a0}, \href {https://ui.adsabs.harvard.edu/abs/1983Natur.301..201D} {301, 201}

\bibitem[\protect\citeauthoryear{{Faramaz} et~al.,}{{Faramaz} et~al.}{2021}]{Faramaz2021}
{Faramaz} V.,  et~al., 2021, \mn@doi [\aj] {10.3847/1538-3881/abf4e0}, \href {https://ui.adsabs.harvard.edu/abs/2021AJ....161..271F} {161, 271}

\bibitem[\protect\citeauthoryear{{Friebe}, {Pearce}  \& {L{\"o}hne}}{{Friebe} et~al.}{2022}]{friebe2022}
{Friebe} M.~F.,  {Pearce} T.~D.,   {L{\"o}hne} T.,  2022, \mn@doi [\mnras] {10.1093/mnras/stac664}, \href {https://ui.adsabs.harvard.edu/abs/2022MNRAS.512.4441F} {512, 4441}

\bibitem[\protect\citeauthoryear{{Geiler}, {Krivov}, {Booth}  \& {L{\"o}hne}}{{Geiler} et~al.}{2019}]{geiler2019}
{Geiler} F.,  {Krivov} A.~V.,  {Booth} M.,   {L{\"o}hne} T.,  2019, \mn@doi [\mnras] {10.1093/mnras/sty3160}, \href {https://ui.adsabs.harvard.edu/abs/2019MNRAS.483..332G} {483, 332}

\bibitem[\protect\citeauthoryear{{Gladman}}{{Gladman}}{1993}]{gladman1993}
{Gladman} B.,  1993, \mn@doi [\icarus] {10.1006/icar.1993.1169}, \href {https://ui.adsabs.harvard.edu/abs/1993Icar..106..247G} {106, 247}

\bibitem[\protect\citeauthoryear{{Gladman} \& {Chan}}{{Gladman} \& {Chan}}{2006}]{Gladman2006}
{Gladman} B.,  {Chan} C.,  2006, \mn@doi [\apjl] {10.1086/505214}, \href {https://ui.adsabs.harvard.edu/abs/2006ApJ...643L.135G} {643, L135}

\bibitem[\protect\citeauthoryear{{Gladman} \& {Volk}}{{Gladman} \& {Volk}}{2021}]{Gladman2021}
{Gladman} B.,  {Volk} K.,  2021, \mn@doi [\araa] {10.1146/annurev-astro-120920-010005}, \href {https://ui.adsabs.harvard.edu/abs/2021ARA&A..59..203G} {59, 203}

\bibitem[\protect\citeauthoryear{{Gomes}, {Levison}, {Tsiganis}  \& {Morbidelli}}{{Gomes} et~al.}{2005}]{Gomes2005}
{Gomes} R.,  {Levison} H.~F.,  {Tsiganis} K.,   {Morbidelli} A.,  2005, \mn@doi [\nat] {10.1038/nature03676}, \href {https://ui.adsabs.harvard.edu/abs/2005Natur.435..466G} {435, 466}

\bibitem[\protect\citeauthoryear{{Gomes}, {Fern{\'a}ndez}, {Gallardo}  \& {Brunini}}{{Gomes} et~al.}{2008}]{Gomes2008}
{Gomes} R.~S.,  {Fern{\'a}ndez} J.~A.,  {Gallardo} T.,   {Brunini} A.,  2008, in {Barucci} M.~A.,  {Boehnhardt} H.,  {Cruikshank} D.~P.,  {Morbidelli} A.,   {Dotson} R.,  eds, , The Solar System Beyond Neptune.
pp 259--273

\bibitem[\protect\citeauthoryear{{Harper}, {Loewenstein}  \& {Davidson}}{{Harper} et~al.}{1984}]{harper1984}
{Harper} D.~A.,  {Loewenstein} R.~F.,   {Davidson} J.~A.,  1984, \mn@doi [\apj] {10.1086/162559}, \href {https://ui.adsabs.harvard.edu/abs/1984ApJ...285..808H} {285, 808}

\bibitem[\protect\citeauthoryear{{Hayashi}}{{Hayashi}}{1981}]{Hayashi1981}
{Hayashi} C.,  1981, \mn@doi [Progress of Theoretical Physics Supplement] {10.1143/PTPS.70.35}, \href {https://ui.adsabs.harvard.edu/abs/1981PThPS..70...35H} {70, 35}

\bibitem[\protect\citeauthoryear{{Huang}, {Gladman}, {Beaudoin}  \& {Zhang}}{{Huang} et~al.}{2022}]{Huang2022}
{Huang} Y.,  {Gladman} B.,  {Beaudoin} M.,   {Zhang} K.,  2022, \mn@doi [\apjl] {10.3847/2041-8213/ac9480}, \href {https://ui.adsabs.harvard.edu/abs/2022ApJ...938L..23H} {938, L23}

\bibitem[\protect\citeauthoryear{{Hughes}, {Duch{\^e}ne}  \& {Matthews}}{{Hughes} et~al.}{2018}]{hughes2018review}
{Hughes} A.~M.,  {Duch{\^e}ne} G.,   {Matthews} B.~C.,  2018, \mn@doi [\araa] {10.1146/annurev-astro-081817-052035}, \href {https://ui.adsabs.harvard.edu/abs/2018ARA&A..56..541H} {56, 541}

\bibitem[\protect\citeauthoryear{{Ida}, {Larwood}  \& {Burkert}}{{Ida} et~al.}{2000a}]{idalarwood2000}
{Ida} S.,  {Larwood} J.,   {Burkert} A.,  2000a, \mn@doi [\apj] {10.1086/308179}, \href {https://ui.adsabs.harvard.edu/abs/2000ApJ...528..351I} {528, 351}

\bibitem[\protect\citeauthoryear{{Ida}, {Bryden}, {Lin}  \& {Tanaka}}{{Ida} et~al.}{2000b}]{ida2000}
{Ida} S.,  {Bryden} G.,  {Lin} D.~N.~C.,   {Tanaka} H.,  2000b, \mn@doi [\apj] {10.1086/308720}, \href {https://ui.adsabs.harvard.edu/abs/2000ApJ...534..428I} {534, 428}

\bibitem[\protect\citeauthoryear{{Imaz Blanco} et~al.,}{{Imaz Blanco} et~al.}{2023}]{ImazBlanco2023}
{Imaz Blanco} A.,  et~al., 2023, \mn@doi [\mnras] {10.1093/mnras/stad1221}, \href {https://ui.adsabs.harvard.edu/abs/2023MNRAS.522.6150I} {522, 6150}

\bibitem[\protect\citeauthoryear{{Kennedy} \& {Wyatt}}{{Kennedy} \& {Wyatt}}{2010}]{kennedy2010}
{Kennedy} G.~M.,  {Wyatt} M.~C.,  2010, \mn@doi [\mnras] {10.1111/j.1365-2966.2010.16528.x}, \href {https://ui.adsabs.harvard.edu/abs/2010MNRAS.405.1253K} {405, 1253}

\bibitem[\protect\citeauthoryear{{Kenyon} \& {Bromley}}{{Kenyon} \& {Bromley}}{2001}]{kenyonbromley2001}
{Kenyon} S.~J.,  {Bromley} B.~C.,  2001, \mn@doi [\aj] {10.1086/318019}, \href {https://ui.adsabs.harvard.edu/abs/2001AJ....121..538K} {121, 538}

\bibitem[\protect\citeauthoryear{{Kenyon} \& {Bromley}}{{Kenyon} \& {Bromley}}{2002}]{kenyonbromley2002}
{Kenyon} S.~J.,  {Bromley} B.~C.,  2002, \mn@doi [\aj] {10.1086/338850}, \href {https://ui.adsabs.harvard.edu/abs/2002AJ....123.1757K} {123, 1757}

\bibitem[\protect\citeauthoryear{{Kenyon} \& {Bromley}}{{Kenyon} \& {Bromley}}{2008}]{kenyon2008}
{Kenyon} S.~J.,  {Bromley} B.~C.,  2008, \mn@doi [\apjs] {10.1086/591794}, \href {https://ui.adsabs.harvard.edu/abs/2008ApJS..179..451K} {179, 451}

\bibitem[\protect\citeauthoryear{{Kenyon} \& {Bromley}}{{Kenyon} \& {Bromley}}{2010}]{kenyonbromley2010selfstirring}
{Kenyon} S.~J.,  {Bromley} B.~C.,  2010, \mn@doi [\apjs] {10.1088/0067-0049/188/1/242}, \href {https://ui.adsabs.harvard.edu/abs/2010ApJS..188..242K} {188, 242}

\bibitem[\protect\citeauthoryear{{Kirsh}, {Duncan}, {Brasser}  \& {Levison}}{{Kirsh} et~al.}{2009}]{kirsh2009}
{Kirsh} D.~R.,  {Duncan} M.,  {Brasser} R.,   {Levison} H.~F.,  2009, \mn@doi [\icarus] {10.1016/j.icarus.2008.05.028}, \href {https://ui.adsabs.harvard.edu/abs/2009Icar..199..197K} {199, 197}

\bibitem[\protect\citeauthoryear{{Kobayashi} \& {Ida}}{{Kobayashi} \& {Ida}}{2001}]{kobayashi2001}
{Kobayashi} H.,  {Ida} S.,  2001, \mn@doi [\icarus] {10.1006/icar.2001.6700}, \href {https://ui.adsabs.harvard.edu/abs/2001Icar..153..416K} {153, 416}

\bibitem[\protect\citeauthoryear{{Krivov}}{{Krivov}}{2010}]{krivov2010review}
{Krivov} A.~V.,  2010, \mn@doi [Research in Astronomy and Astrophysics] {10.1088/1674-4527/10/5/001}, \href {https://ui.adsabs.harvard.edu/abs/2010RAA....10..383K} {10, 383}

\bibitem[\protect\citeauthoryear{{Krivov} \& {Booth}}{{Krivov} \& {Booth}}{2018}]{krivov2018}
{Krivov} A.~V.,  {Booth} M.,  2018, \mn@doi [\mnras] {10.1093/mnras/sty1607}, \href {https://ui.adsabs.harvard.edu/abs/2018MNRAS.479.3300K} {479, 3300}

\bibitem[\protect\citeauthoryear{{Krivov} \& {Wyatt}}{{Krivov} \& {Wyatt}}{2021}]{krivov2021}
{Krivov} A.~V.,  {Wyatt} M.~C.,  2021, \mn@doi [\mnras] {10.1093/mnras/staa2385}, \href {https://ui.adsabs.harvard.edu/abs/2021MNRAS.500..718K} {500, 718}

\bibitem[\protect\citeauthoryear{{Krivov}, {Srem{\v{c}}evi{\'c}}  \& {Spahn}}{{Krivov} et~al.}{2005}]{krivov2005}
{Krivov} A.~V.,  {Srem{\v{c}}evi{\'c}} M.,   {Spahn} F.,  2005, \mn@doi [\icarus] {10.1016/j.icarus.2004.10.003}, \href {https://ui.adsabs.harvard.edu/abs/2005Icar..174..105K} {174, 105}

\bibitem[\protect\citeauthoryear{{Krivov}, {Queck}, {L{\"o}hne}  \& {Srem{\v{c}}evi{\'c}}}{{Krivov} et~al.}{2007}]{Krivov2007}
{Krivov} A.~V.,  {Queck} M.,  {L{\"o}hne} T.,   {Srem{\v{c}}evi{\'c}} M.,  2007, \mn@doi [\aap] {10.1051/0004-6361:20065584}, \href {https://ui.adsabs.harvard.edu/abs/2007A&A...462..199K} {462, 199}

\bibitem[\protect\citeauthoryear{{Krivov}, {Ide}, {L{\"o}hne}, {Johansen}  \& {Blum}}{{Krivov} et~al.}{2018}]{krivovqdstar2018}
{Krivov} A.~V.,  {Ide} A.,  {L{\"o}hne} T.,  {Johansen} A.,   {Blum} J.,  2018, \mn@doi [\mnras] {10.1093/mnras/stx2932}, \href {https://ui.adsabs.harvard.edu/abs/2018MNRAS.474.2564K} {474, 2564}

\bibitem[\protect\citeauthoryear{{Malhotra}}{{Malhotra}}{2019}]{Malhotra2019}
{Malhotra} R.,  2019, \mn@doi [Geoscience Letters] {10.1186/s40562-019-0142-2}, \href {https://ui.adsabs.harvard.edu/abs/2019GSL.....6...12M} {6, 12}

\bibitem[\protect\citeauthoryear{{Malhotra}, {Castro-Cisneros}, {Fitzgerald}, {Rosengren}, {Ross}, {Todorovic}  \& {Wu}}{{Malhotra} et~al.}{2021}]{Malhotra2021}
{Malhotra} R.,  {Castro-Cisneros} J.~D.,  {Fitzgerald} J.,  {Rosengren} A.~J.,  {Ross} S.,  {Todorovic} N.,   {Wu} D.,  2021, in AAS/Division of Dynamical Astronomy Meeting. p. 305.04

\bibitem[\protect\citeauthoryear{{Marois}, {Macintosh}, {Barman}, {Zuckerman}, {Song}, {Patience}, {Lafreni{\`e}re}  \& {Doyon}}{{Marois} et~al.}{2008}]{Marois2008}
{Marois} C.,  {Macintosh} B.,  {Barman} T.,  {Zuckerman} B.,  {Song} I.,  {Patience} J.,  {Lafreni{\`e}re} D.,   {Doyon} R.,  2008, \mn@doi [Science] {10.1126/science.1166585}, \href {https://ui.adsabs.harvard.edu/abs/2008Sci...322.1348M} {322, 1348}

\bibitem[\protect\citeauthoryear{{Matr{\`a}}, {Marino}, {Kennedy}, {Wyatt}, {{\"O}berg}  \& {Wilner}}{{Matr{\`a}} et~al.}{2018}]{matra2018}
{Matr{\`a}} L.,  {Marino} S.,  {Kennedy} G.~M.,  {Wyatt} M.~C.,  {{\"O}berg} K.~I.,   {Wilner} D.~J.,  2018, \mn@doi [\apj] {10.3847/1538-4357/aabcc4}, \href {https://ui.adsabs.harvard.edu/abs/2018ApJ...859...72M} {859, 72}

\bibitem[\protect\citeauthoryear{{Matr{\`a}}, {Wyatt}, {Wilner}, {Dent}, {Marino}, {Kennedy}  \& {Milli}}{{Matr{\`a}} et~al.}{2019}]{matra2019}
{Matr{\`a}} L.,  {Wyatt} M.~C.,  {Wilner} D.~J.,  {Dent} W.~R.~F.,  {Marino} S.,  {Kennedy} G.~M.,   {Milli} J.,  2019, \mn@doi [\aj] {10.3847/1538-3881/ab06c0}, \href {https://ui.adsabs.harvard.edu/abs/2019AJ....157..135M} {157, 135}

\bibitem[\protect\citeauthoryear{{Matthews}, {Krivov}, {Wyatt}, {Bryden}  \& {Eiroa}}{{Matthews} et~al.}{2014}]{matthews2014review}
{Matthews} B.~C.,  {Krivov} A.~V.,  {Wyatt} M.~C.,  {Bryden} G.,   {Eiroa} C.,  2014, in {Beuther} H.,  {Klessen} R.~S.,  {Dullemond} C.~P.,   {Henning} T.,  eds, Protostars and Planets VI. p.~521 (\mn@eprint {arXiv} {1401.0743}), \mn@doi{10.2458/azu_uapress_9780816531240-ch023}

\bibitem[\protect\citeauthoryear{{Morbidelli}, {Levison}, {Tsiganis}  \& {Gomes}}{{Morbidelli} et~al.}{2005}]{Morbidelli2005}
{Morbidelli} A.,  {Levison} H.~F.,  {Tsiganis} K.,   {Gomes} R.,  2005, \mn@doi [\nat] {10.1038/nature03540}, \href {https://ui.adsabs.harvard.edu/abs/2005Natur.435..462M} {435, 462}

\bibitem[\protect\citeauthoryear{{Mu{\~n}oz-Guti{\'e}rrez}, {Pichardo}, {Reyes-Ruiz}  \& {Peimbert}}{{Mu{\~n}oz-Guti{\'e}rrez} et~al.}{2015}]{munoz2015dpdiskexcitation}
{Mu{\~n}oz-Guti{\'e}rrez} M.~A.,  {Pichardo} B.,  {Reyes-Ruiz} M.,   {Peimbert} A.,  2015, \mn@doi [\apjl] {10.1088/2041-8205/811/2/L21}, \href {https://ui.adsabs.harvard.edu/abs/2015ApJ...811L..21M} {811, L21}

\bibitem[\protect\citeauthoryear{{Mu{\~n}oz-Guti{\'e}rrez}, {Marshall}  \& {Peimbert}}{{Mu{\~n}oz-Guti{\'e}rrez} et~al.}{2023}]{munozmixedstirring2023}
{Mu{\~n}oz-Guti{\'e}rrez} M.~A.,  {Marshall} J.~P.,   {Peimbert} A.,  2023, \mn@doi [\mnras] {10.1093/mnras/stad218}, \href {https://ui.adsabs.harvard.edu/abs/2023MNRAS.tmp..232M} {}

\bibitem[\protect\citeauthoryear{Murray \& Dermott}{Murray \& Dermott}{1999}]{murray1999}
Murray C.~D.,  Dermott S.~F.,  1999, Solar System Dynamics.
Cambridge University Press, \mn@doi{10.1017/CBO9781139174817}

\bibitem[\protect\citeauthoryear{{Mustill} \& {Wyatt}}{{Mustill} \& {Wyatt}}{2009}]{mustill2009}
{Mustill} A.~J.,  {Wyatt} M.~C.,  2009, \mn@doi [\mnras] {10.1111/j.1365-2966.2009.15360.x}, \href {https://ui.adsabs.harvard.edu/abs/2009MNRAS.399.1403M} {399, 1403}

\bibitem[\protect\citeauthoryear{{Najita}, {Kenyon}  \& {Bromley}}{{Najita} et~al.}{2022}]{najita2022}
{Najita} J.~R.,  {Kenyon} S.~J.,   {Bromley} B.~C.,  2022, \mn@doi [\apj] {10.3847/1538-4357/ac37b6}, \href {https://ui.adsabs.harvard.edu/abs/2022ApJ...925...45N} {925, 45}

\bibitem[\protect\citeauthoryear{{Nesvorn{\'y}}}{{Nesvorn{\'y}}}{2018}]{Nesvorny2018}
{Nesvorn{\'y}} D.,  2018, \mn@doi [\araa] {10.1146/annurev-astro-081817-052028}, \href {https://ui.adsabs.harvard.edu/abs/2018ARA&A..56..137N} {56, 137}

\bibitem[\protect\citeauthoryear{{Pearce} \& {Wyatt}}{{Pearce} \& {Wyatt}}{2014}]{Pearce2014}
{Pearce} T.~D.,  {Wyatt} M.~C.,  2014, \mn@doi [\mnras] {10.1093/mnras/stu1302}, \href {https://ui.adsabs.harvard.edu/abs/2014MNRAS.443.2541P} {443, 2541}

\bibitem[\protect\citeauthoryear{{Pearce} \& {Wyatt}}{{Pearce} \& {Wyatt}}{2015}]{Pearce2015}
{Pearce} T.~D.,  {Wyatt} M.~C.,  2015, \mn@doi [\mnras] {10.1093/mnras/stv1847}, \href {https://ui.adsabs.harvard.edu/abs/2015MNRAS.453.3329P} {453, 3329}

\bibitem[\protect\citeauthoryear{{Pearce}, {Beust}, {Faramaz}, {Booth}, {Krivov}, {L{\"o}hne}  \& {Poblete}}{{Pearce} et~al.}{2021}]{Pearce2021}
{Pearce} T.~D.,  {Beust} H.,  {Faramaz} V.,  {Booth} M.,  {Krivov} A.~V.,  {L{\"o}hne} T.,   {Poblete} P.~P.,  2021, \mn@doi [\mnras] {10.1093/mnras/stab760}, \href {https://ui.adsabs.harvard.edu/abs/2021MNRAS.503.4767P} {503, 4767}

\bibitem[\protect\citeauthoryear{{Pearce} et~al.,}{{Pearce} et~al.}{2022}]{pearceispy2022}
{Pearce} T.~D.,  et~al., 2022, \mn@doi [\aap] {10.1051/0004-6361/202142720}, \href {https://ui.adsabs.harvard.edu/abs/2022A&A...659A.135P} {659, A135}

\bibitem[\protect\citeauthoryear{{Pearce} et~al.,}{{Pearce} et~al.}{2023}]{Pearce2023}
{Pearce} T.~D.,  et~al., 2023, \mn@doi [arXiv e-prints] {10.48550/arXiv.2311.04265}, \href {https://ui.adsabs.harvard.edu/abs/2023arXiv231104265P} {p. arXiv:2311.04265}

\bibitem[\protect\citeauthoryear{{Petrovich}, {Malhotra}  \& {Tremaine}}{{Petrovich} et~al.}{2013}]{Petrovich2013}
{Petrovich} C.,  {Malhotra} R.,   {Tremaine} S.,  2013, \mn@doi [\apj] {10.1088/0004-637X/770/1/24}, \href {https://ui.adsabs.harvard.edu/abs/2013ApJ...770...24P} {770, 24}

\bibitem[\protect\citeauthoryear{{Rein} \& {Liu}}{{Rein} \& {Liu}}{2012}]{rebound}
{Rein} H.,  {Liu} S.~F.,  2012, \mn@doi [\aap] {10.1051/0004-6361/201118085}, \href {https://ui.adsabs.harvard.edu/abs/2012A&A...537A.128R} {537, A128}

\bibitem[\protect\citeauthoryear{{Rein} et~al.,}{{Rein} et~al.}{2019}]{reboundmercurius}
{Rein} H.,  et~al., 2019, \mn@doi [\mnras] {10.1093/mnras/stz769}, \href {https://ui.adsabs.harvard.edu/abs/2019MNRAS.485.5490R} {485, 5490}

\bibitem[\protect\citeauthoryear{{Smith} \& {Terrile}}{{Smith} \& {Terrile}}{1984}]{smith1984scattered}
{Smith} B.~A.,  {Terrile} R.~J.,  1984, \mn@doi [Science] {10.1126/science.226.4681.1421}, \href {https://ui.adsabs.harvard.edu/abs/1984Sci...226.1421S} {226, 1421}

\bibitem[\protect\citeauthoryear{{Tabeshian} \& {Wiegert}}{{Tabeshian} \& {Wiegert}}{2016}]{Tabeshian2016}
{Tabeshian} M.,  {Wiegert} P.~A.,  2016, \mn@doi [\apj] {10.3847/0004-637X/818/2/159}, \href {https://ui.adsabs.harvard.edu/abs/2016ApJ...818..159T} {818, 159}

\bibitem[\protect\citeauthoryear{{Tremaine}}{{Tremaine}}{1993}]{tremaine1993}
{Tremaine} S.,  1993, in {Phillips} J.~A.,  {Thorsett} S.~E.,   {Kulkarni} S.~R.,  eds,  Astronomical Society of the Pacific Conference Series Vol. 36, Planets Around Pulsars. pp 335--344

\bibitem[\protect\citeauthoryear{{Tsiganis}, {Gomes}, {Morbidelli}  \& {Levison}}{{Tsiganis} et~al.}{2005}]{Tsiganis2005}
{Tsiganis} K.,  {Gomes} R.,  {Morbidelli} A.,   {Levison} H.~F.,  2005, \mn@doi [\nat] {10.1038/nature03539}, \href {https://ui.adsabs.harvard.edu/abs/2005Natur.435..459T} {435, 459}

\bibitem[\protect\citeauthoryear{{Walmswell}, {Clarke}  \& {Cossins}}{{Walmswell} et~al.}{2013}]{walmswell2013}
{Walmswell} J.,  {Clarke} C.,   {Cossins} P.,  2013, \mn@doi [\mnras] {10.1093/mnras/stt314}, \href {https://ui.adsabs.harvard.edu/abs/2013MNRAS.431.1903W} {431, 1903}

\bibitem[\protect\citeauthoryear{{Weidenschilling}}{{Weidenschilling}}{1977}]{Weidenschilling1977}
{Weidenschilling} S.~J.,  1977, \mn@doi [\apss] {10.1007/BF00642464}, \href {https://ui.adsabs.harvard.edu/abs/1977Ap&SS..51..153W} {51, 153}

\bibitem[\protect\citeauthoryear{{Weissman}}{{Weissman}}{1984}]{weissman1984}
{Weissman} P.~R.,  1984, \mn@doi [Science] {10.1126/science.224.4652.987}, \href {https://ui.adsabs.harvard.edu/abs/1984Sci...224..987W} {224, 987}

\bibitem[\protect\citeauthoryear{{Whitmire}, {Matese}, {Criswell}  \& {Mikkola}}{{Whitmire} et~al.}{1998}]{whitmire1998}
{Whitmire} D.~P.,  {Matese} J.~J.,  {Criswell} L.,   {Mikkola} S.,  1998, \mn@doi [\icarus] {10.1006/icar.1998.5900}, \href {https://ui.adsabs.harvard.edu/abs/1998Icar..132..196W} {132, 196}

\bibitem[\protect\citeauthoryear{{Wyatt}}{{Wyatt}}{2008}]{wyatt2008}
{Wyatt} M.~C.,  2008, \mn@doi [\araa] {10.1146/annurev.astro.45.051806.110525}, \href {https://ui.adsabs.harvard.edu/abs/2008ARA&A..46..339W} {46, 339}

\bibitem[\protect\citeauthoryear{{Wyatt}}{{Wyatt}}{2020}]{wyatt2020}
{Wyatt} M.,  2020, in {Prialnik} D.,  {Barucci} M.~A.,   {Young} L.,  eds, , The Trans-Neptunian Solar System.
pp 351--376, \mn@doi{10.1016/B978-0-12-816490-7.00016-3}

\bibitem[\protect\citeauthoryear{{Wyatt} \& {Dent}}{{Wyatt} \& {Dent}}{2002}]{wyatt2002}
{Wyatt} M.~C.,  {Dent} W.~R.~F.,  2002, \mn@doi [\mnras] {10.1046/j.1365-8711.2002.05533.x}, \href {https://ui.adsabs.harvard.edu/abs/2002MNRAS.334..589W} {334, 589}

\bibitem[\protect\citeauthoryear{{Wyatt}, {Dermott}, {Telesco}, {Fisher}, {Grogan}, {Holmes}  \& {Pi{\~n}a}}{{Wyatt} et~al.}{1999}]{wyatt1999}
{Wyatt} M.~C.,  {Dermott} S.~F.,  {Telesco} C.~M.,  {Fisher} R.~S.,  {Grogan} K.,  {Holmes} E.~K.,   {Pi{\~n}a} R.~K.,  1999, \mn@doi [\apj] {10.1086/308093}, \href {https://ui.adsabs.harvard.edu/abs/1999ApJ...527..918W} {527, 918}

\makeatother
\end{thebibliography}

%%%%%%%%%%%%%%%%%%%%%%%%%%%%%%%%%%%%%%%%%%%%%%%%%%

%%%%%%%%%%%%%%%%% APPENDICES %%%%%%%%%%%%%%%%%%%%%

\appendix

%#############################################################################################################################
\section{Stirring analyses}
\label{sec: stirring_analysis_appendix}

This appendix provides more details of our analytic and numerical stirring analyses.

\subsection{Catastrophic collisions}

We consider dust released through catastrophic collisions between planetesimals. For a collision to be catastrophic, its kinetic energy needs to be greater than the critical fragmentation energy, $\QDStar$, such that the largest remnant after the collision has half the original mass of the planetesimal that broke apart (\citealt{wyatt2002}).  Approximating the debris particles as basalt, $\QDStar$ is given by

\begin{equation}
    \QDStar(s,v_{\rm coll}) = 9.13~{\rm J~kg^{-1}} \left(\frac{v_{\rm coll}}{\rm m~s^{-1}}\right)^{1/2} \left[ \left(\frac{s}{\rm m}\right)^{\rm -0.36} + \left(\frac{s}{\rm km}\right)^{\rm 1.4}\right],
    \label{eq: qdstar}
\end{equation}

\noindent where $v_{\rm coll}$ is the collisional speed and $s$ is the debris-body radius (see Equation 1 in \citealt{krivovqdstar2018}). 

 This $\QDStar$ prescription results in two regimes in the shape of a ``v''.  For basalt, the minimum $\QDStar$ is for debris with size 110~m (\citealt{pearceispy2022}), meaning that bodies of this size are easiest to break apart. Smaller debris is considered to be in the strength regime; in this regime, $\QDStar$ is approximately equal to the shattering strength (\citealt{wyatt2002}). Bodies larger than 110~m are considered to be in the gravity regime, where additional energy is needed to overcome the planetesimal's gravity and break it apart.

Catastrophic collisions will occur at the onset of the collisional cascade between particles of similar sizes (\citealt{pearceispy2022}).  From this, we can define the fragmentation speed (dependant on debris size), under the assumption that the material is basalt (see \citealt{krivov2005}): 

\begin{equation}
    v_{\rm frag}(s) = 17.5\left[\left(\frac{s}{\rm m}\right)^{-0.36} + \left(\frac{s}{\rm km}\right)^{1.4}\right]^{2/3}~\rm {m/s}.
    \label{eq: fragmentation_velocity_app}
\end{equation}

\noindent Note that the equation for fragmentation speed may vary for a given size when considering materials other than basalt.

%-----------------------------------------------------------------------------------------------------------------------------
\subsection{Simple analytic stirring analysis}
\label{sec: simplified_stirring_analysis_appendix}

In this section we detail the derivation of Equations \ref{eq: unstirred_ecc} to \ref{eq: final_unstirred_ecc}, which provide a simple approximation of the minimum debris eccentricity required for stirring.

\begin{figure}
    \centering
    \includegraphics[width=8cm]{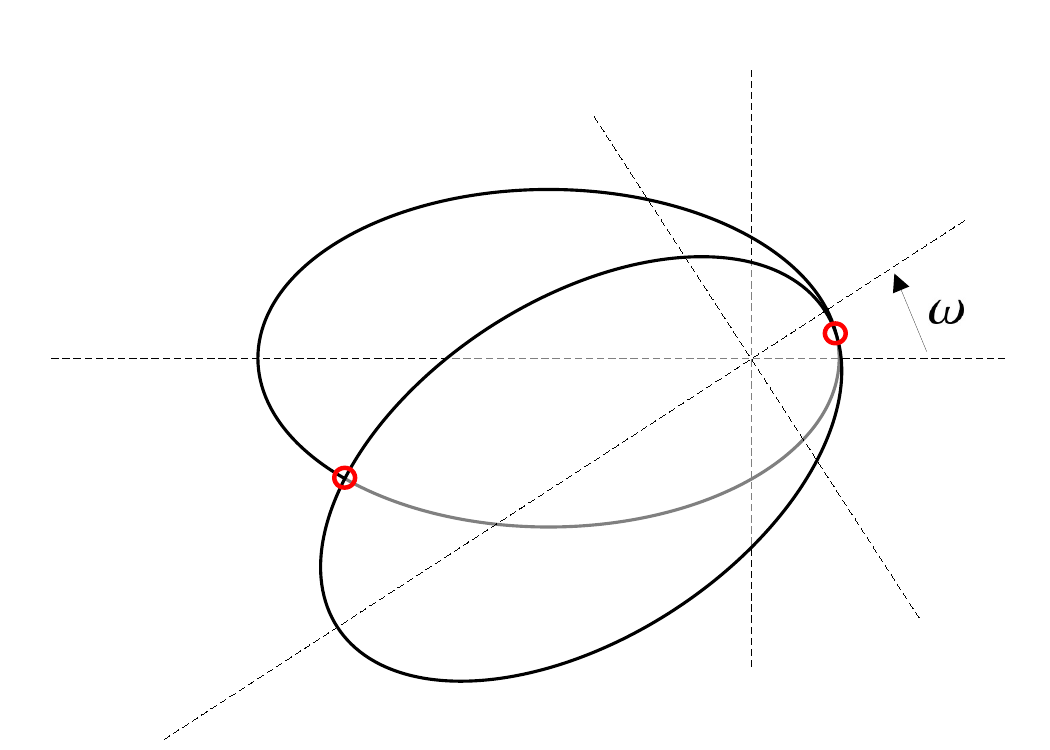}
    \caption{Diagram depicting the orbits of two similar debris bodies, as considered in the simple stirring analysis in Section \ref{subsec: simpleStirringCriterion} and Appendix \ref{sec: simplified_stirring_analysis_appendix}.  It is assumed that the debris particles have the same semimajor axis and eccentricity, and are rotationally offset by angle $\omega$.  There are two points where the particles collide, located at the red circles.}
    \label{fig: simple_velocity_diagram}
\end{figure}

Figure \ref{fig: simple_velocity_diagram} depicts two debris particles with crossing orbits, with the assumptions of co-planar particles of equal size, semimajor axis and eccentricity.  We aim to solve for the collisional speeds at the points where the orbits cross, in terms of the rotational angle $\omega$. Since the orbits (defined with subscripts 1 and 2 for the two bodies in subsequent equations) are rotated from one another, they will have differing true anomalies, \textit{f}, at the intersection points. At the intersecting points, the orbits will have the same $x$ and $y$ positions; hence we can equate ${x_1=x_2}$ and ${y_1=y_2}$ at these locations. The distances $x$ and $y$ are given by equations $x = r\cos{f}$ and $y = r\sin{f}$, with $r = a(1-e^{2})(1+e \cos{f})^{-1}$ (Equations 2.21 and 2.20 in \citealt{murray1999}, respectively). This yields

\begin{equation}
    \centering
    \cos{f_{\rm 1}} + e\cos{f_{\rm 1}}\cos{f_{\rm 2}} = \cos{(f_{\rm 2}+\omega)} + e\cos{f_{\rm 1}}\cos{(f_{\rm 2}+\omega)},
    \label{eq: initial_vel_simplification_partial_1}
\end{equation}

\noindent and

\begin{equation}
    \centering
    \sin{f_{\rm 1}} + e\sin{f_{\rm 1}}\cos{f_{\rm 2}} = \sin{(f_{\rm 2}+\omega)} + e\cos{f_{\rm 1}}\sin{(f_{\rm 2}+\omega)}.
    \label{eq: initial_vel_simplification_partial_2}
\end{equation}

\noindent The two solutions for the two intersection points are

\begin{equation}
        f_{\rm 1} = \frac{\omega}{2}, \pi + \frac{\omega}{2},
        \label{eq: vel_solution_f_1}
\end{equation}

\noindent and

\begin{equation}
    \centering
    f_{\rm 2} = -\frac{\omega}{2}, \pi - \frac{\omega}{2},
    \label{eq: vel_solution_f_2}
\end{equation}

\noindent where all angles are in radians.

With the true anomalies at the intersection points calculated, it is possible to solve for the collisional speed of these debris particles.  The collisional speed is

\begin{equation}
    v_{\rm coll} = \sqrt{(v_{\rm x,1} - v_{\rm x,2})^2 + (v_{\rm y,1} - v_{\rm y,2})^2},
    \label{eq: initial_collision_vel}
\end{equation}

\noindent where \textit{$v_{\rm x,i}$} and \textit{$v_{\rm y,i}$} are the velocities in the \textit{x} and \textit{y} directions for each particle.  They are

\begin{equation}
    \begin{aligned}
        v_{\rm x,1} &= -A \sin{f_{\rm 1}}\\
        v_{\rm y,1} &= A(e+\cos{f_{\rm 1}}),
        \label{eq: velocity_eq_particle_1}
    \end{aligned}
\end{equation}

\noindent and

\begin{equation}
    \begin{aligned}
        v_{\rm x,2} &= -A \sin{f_{\rm 2}}\cos{\omega} - A(e+\cos{f_{\rm 2}})\sin{\omega}\\
        v_{\rm y,2} &= -A\sin{f_{\rm 2}}\sin{\omega} + A(e+\cos{f_{\rm 2}})\cos{\omega},
        \label{eq: velocity_eq_particle_2}
    \end{aligned}
\end{equation}

\noindent where the velocities for particle 2 have been rotated by angle $\omega$, and the constant \textit{A} is given by $A = \left(G\mStar\right)^{1/2}\left[{a(1-e^2)}\right]^{-1/2}$.  These velocities are given by \mbox{Equation 2.36} in \cite{murray1999}, with the pre-factor \textit{A} being the same but in a different form.

To determine whether these particles are considered stirred, we solve for the collisional speed at the locations where the orbits cross.  Substituting Equations \ref{eq: velocity_eq_particle_1} and \ref{eq: velocity_eq_particle_2} into Equation \ref{eq: initial_collision_vel} results in the collisional speed equation:

\begin{equation}
    v_{\rm coll,simple} = e \sqrt{\frac{2 G\mStar(1-\cos{\omega})}{a(1-e^2)}}.
    \label{eq: final_collision_vel}
\end{equation}

\noindent Note that with the assumption that both debris particles have the same semimajor axis and eccentricity, and are offset by angle $\omega$, both solutions for $f_1$ and $f_2$ result in the same simplified collisional speed equation. Therefore, the solution of ${f_1 = \omega /2, f_2 = -\omega/2}$ is used.

\begin{figure}
    \centering
    \includegraphics[width=8cm]{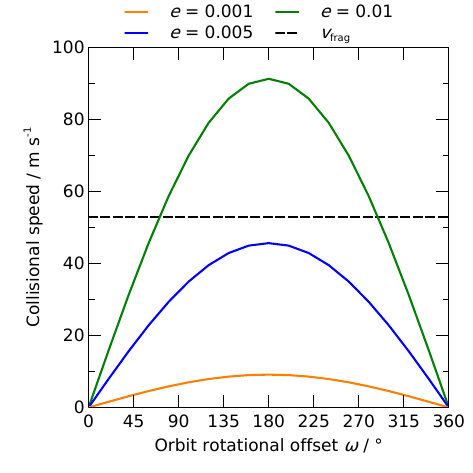}
    \caption{Collisional speed of two coplanar particles with identical semimajor axes and eccentricities, at varying eccentricities and orbit-rotational offsets.  The collisional speed is greater the more anti-aligned the two colliding debris particle orbits are, and also for higher eccentricities.  The fragmentation speed is plotted as a guide of the speed needed to consider particles stirred.  To make the plot, values of ${a = 42.5\au}$, ${\mStar = 1\mSun}$, and ${s = 1\cm}$ are used.}
    \label{fig: velocity_orientation}
\end{figure}

To consider the eccentricity below which debris particles will certainly not be stirred, Equations \ref{eq: final_collision_vel} and \ref{eq: fragmentation_velocity_app} are compared to find where the collisional speed is less than the fragmentation speed. We are specifically looking for the boundary below which it is certain the debris particles will be unstirred (for the given setup), so to minimize the eccentricity the particles' orbits are considered to be anti-aligned, i.e. $\omega = 180\degree$ (see Figure \ref{fig: velocity_orientation}).  Doing so, the unstirred eccentricity boundary can be defined as:

\begin{equation}
    e_{\rm unstirred} < \frac{v_{\rm frag}}{2} \sqrt{\frac{\aDeb}{G \mStar}}.
    \label{eq: unstirred_ecc_appendix}
\end{equation}

\noindent We now plot the collisional speed against semimajor axis, eccentricity, grain size, and orbital rotation $\omega$ to see how the fragmentation and collisional speeds vary.

Figure \ref{fig: velocity_orientation} shows how the collisional speed changes with orbit-rotational offset $\omega$. The collisional speed increases the more anti-aligned the orbits are, reaching a maximum at 180\degree.  As seen in the figure, higher eccentricities result in a higher collisional speed.  It can also be seen that it is possible for the collisional speed to be larger than the fragmentation speed even if the orbits are not fully anti-aligned, as long as the eccentricity of the debris particles is sufficiently high.

\begin{figure}
    \centering
    \includegraphics[width=8cm]{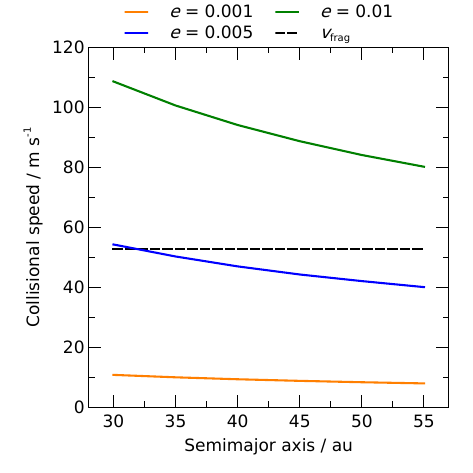}
    \caption{Collisional speed of two coplanar particles with identical semimajor axes and eccentricities, at varying eccentricities and semimajor axes.  The collisional speed is greater at higher eccentricities, but decreases the further away the particles are from the star.  The fragmentation speed is plotted as a guide of the speed needed to consider particles stirred.  To make the plot, values of ${\omega = 180\degree}$, ${\mStar = 1\mSun}$, and ${s = 1\cm}$ are used.}
    \label{fig: velocity_semimajor_axis}
\end{figure}

Figure \ref{fig: velocity_semimajor_axis} shows how the collisional speed varies compared to the semimajor axis for given eccentricities.  The collisional speed decreases with increasing semimajor axis, and larger eccentricities provide a greater collisional speed.

\begin{figure}
    \centering
    \includegraphics[width=8cm]{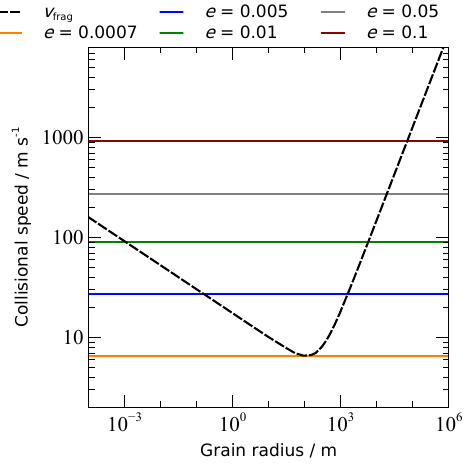}
    \caption{Collisional speed of two coplanar particles with identical semimajor axes and eccentricities, at varying eccentricities and particle sizes.  The fragmentation speed (given by the dashed black line) changes with particle size. This is calculated assuming the debris particles are basalt, which has a minimum fragmentation speed at a grain size of ${110\m}$. To make the plot, values of ${\omega = 180\degree}$, ${a = 42.5\au}$,  and ${\mStar = 1\mSun}$ are used.}
    \label{fig: velocity_particle_size}
\end{figure}

The collisional speed can also be compared to the fragmentation speed by considering the grain size, as on Figure \ref{fig: velocity_particle_size}.  The collisional speed is constant with grain size for a given eccentricity, however the fragmentation speed varies with grain size. For basalt, the weakest debris particles would have a radius of 110~m, which can be seen on Figure \ref{fig: velocity_particle_size}.  The fragmentation speed increases in either direction from the grain size of 110~m.  As also seen on the plot, there is a minimum particle eccentricity needed so that the collisional speed can be greater than the fragmentation speed at the weakest point.  This minimum eccentricity increases when considering debris particles smaller or larger than 110~m.

%-----------------------------------------------------------------------------------------------------------------------------
\subsection{Numerical stirring analysis}
\label{sec: general_stirring_analysis_appendix}

In Appendix \ref{sec: simplified_stirring_analysis_appendix} we derived a rough criterion for the minimum eccentricity required for debris to be stirred. However, this is not sufficient to define whether debris is actually stirred, because it assumes that two bodies are on identical, optimally misaligned orbits. To properly assess stirring in our N-body simulations, we create a numerical code to assess the actual collision speeds. The workings of the code are described in this section.

%-----------------------------------------------------------------------------------------------------------------------------
\subsubsection{Generalised collision speed}

For the case where two coplanar particles have different orbits with any orientation, we can once again solve for the collisional speed by considering the points where the two orbits intersect.  The solutions to the angles of the collision points are given by Equation 4 in \cite{whitmire1998}:

\begin{equation}
    \cos{\theta} = \frac{-AB \pm C\sqrt{C^2 + B^2 - A^2}}{B^2 + C^2},
    \label{eq: crossing_orbit_angle_solution}
\end{equation}

\noindent where

\begin{equation}
    \begin{aligned}
        A = & \; p_{\rm 2} - p_{\rm 1} \\
        B = & \; e_{\rm 1}p_{\rm 2}\cos(\Delta \varpi) - e_{\rm 2}p_{\rm 1} \\
        C = & \; e_{\rm 1}p_{\rm 2}\sin(\Delta \varpi),
        \label{eq: crossing_orbit_constants}
    \end{aligned}
\end{equation}

\noindent where the subscripts again correspond to debris particles 1 and 2.  In these equations, $p_{\rm i}$ is the semilatus rectum of the particle, and $\Delta \varpi$ is the difference in the two particles' longitudes of pericentre.  The orbits will cross if there are real solutions to Equation \ref{eq: crossing_orbit_angle_solution}, which occurs when the discriminant is positive, i.e. if

\begin{equation}
    C^2 + B^2 - A^2 \ge 0.
    \label{eq: doOrbitsIntersect}
\end{equation}

If there are real solutions, then the relative velocities at the intersection points can be calculated as

\begin{equation}
    \begin{aligned}
        v_{\rm coll} =&\left\{\left[e_{\rm 1}\sin{\left(\theta_{\pm}-\Delta \varpi\right)} \left(\frac{p_{\rm 1}}{\rm au}\right)^{-1/2} - e_{\rm 2}\sin{\left(\theta_{\pm}\right)}\left(\frac{p_{\rm 2}}{\rm au}\right)^{-1/2}\right]^2 \right. \\
        &\left.+ \left[\left(\frac{p_{\rm 1}}{\rm au}\right)^{1/2}\left(\frac{r_{\rm \pm}}{\rm au}\right)^{-1} - \left(\frac{p_{\rm 2}}{\rm au}\right)^{1/2}\left(\frac{r_{\rm \pm}}{\rm au}\right)^{-1}\right]^{2}\right\}^{1/2} \\
        &\times~29.78~{\rm km/s} \left( \frac{\mStar}{\mSun}\right)^{1/2},
    \end{aligned}
    \label{eq: relative_velocities}
\end{equation}

\noindent which is modified from Equation 7 in \cite{whitmire1998} to include the orbital speed mentioned in their text. In this equation $r_{\rm \pm}$ and $\theta_{\rm \pm}$ define the locations of the intersection points, where

\begin{equation}
    r_{\rm \pm} = \frac{p_{\rm 2}}{1+e_{\rm 2} \cos{\theta_{\rm \pm}}}.
    \label{eq: r_pm}
\end{equation}

%-----------------------------------------------------------------------------------------------------------------------------
\subsubsection{Numerical implementation}

We want to calculate the level of stirring in a \textsc{rebound} N-body simulation.  To do so, we develop a bespoke \textsc{python} script capable of performing this calculation. The code considers pairs of debris particles, and checks whether their orbits intersect using \mbox{Equation \ref{eq: doOrbitsIntersect}}. If they intersect, then the code calculates their collision speed using Equation \ref{eq: relative_velocities}; if their collision speed is greater than the fragmentation speed for centimetre-sized debris (Equation \ref{eq: fragmentation_velocity_app}, where ${v_{\rm frag}(1~{\rm cm}) \approx 52.8\mPerS}$), then the two particles are considered stirred. In this section we describe the numerical implementation, which we apply to each of our {\sc rebound} simulations once those simulation have finished.

To check the amount of stirring in an N-body simulation, each particle in the simulated debris disk needs to be checked against other disk particles, to determine whether the conditions mentioned above are fulfilled. This is done at various times throughout the disk's evolution, to account debris orbits changing over time. To avoid unrealistic scenarios, such as high-eccentricity particles that get scattered from the inner edge of the disk colliding with those at the outer edge before colliding with a nearby particle, or one debris particle colliding with many other particles, some criteria are implemented when comparing pairs of debris particles:

\begin{itemize}
    \item If two debris particles collide and are considered stirred, they cannot collide with any other particles, either at this time or at any other time in the simulation (i.e. once a particle is stirred, it is assumed to have been collisionally destroyed).
    \item Debris-particle orbits must be bound (i.e. $e < 1$).
    \item Debris particles' semimajor axes must have been within a certain range of each other at the start of the simulation (to avoid scattered particles moving through the disk at high eccentricities, and colliding with everything).
    \item Debris particles' semimajor axes must be within a certain range of each other at the current time.
    \item Debris particles' semimajor axes at the current time must be within a certain range of their initial values (particles whose semimajor axes have changed considerably have probably been scattered, and are therefore not considered).
    \item Debris particles' inclinations must be within a certain range of each other at the current time (orbits with high mutual inclination may satisfy the two-dimensional intersection criterion from Equation \ref{eq: doOrbitsIntersect}, but would not intersect in reality).
\end{itemize}

\noindent Only pairs of debris particles satisfying all of the above criteria are considered in the collision-speed analysis.

Different values are tested to determine an appropriate set of numerical values for the above conditions. These parameters are ``Particle Near Range'' (PNR), which defines how close the semimajor axes of a particle pair should be (at both the initial and current times), ``Particle Initial Range'' (PIR), which defines how close each particle's semimajor axis should be to its initial value, and ``Inclination Range'' (IR), which defines the maximum mutual inclination between particle pairs. We test multiple values for each of these numerical parameters, across multiple simulations, to ensure that they provide realistic and expected results.  Table \ref{tab:stirring_params} shows the results of one such test, demonstrating how the measured stirring level in the projectile simulation on Figure \ref{fig: projectileVsPltOnlySims} changes as the numerical parameters are varied. Based on these tests, we select values of ${{\rm PNR} = 1\percent}$, ${{\rm PIR} = 1\percent}$, and ${{\rm IR} = 1\degree}$. These values are chosen as a compromise between accuracy and speed.

\begin{table}
    \centering
    \caption{Variation of the measured stirring level for the projectile simulation on Figure \ref{fig: projectileVsPltOnlySims}, as a function of numerical parameters. The simulation was checked every 5 snapshots, and particles were considered scattered if their semimajor axes changed by more than ${20\percent}$. The values shown are the fraction of debris that is stirred but not scattered.}
    \label{tab:stirring_params}
    \begin{tabular}{cccc} % four columns, alignment for each
        \hline
        PNR / & PIR / & IR / & Fraction of disk stirred / \\
        per cent & per cent & \degree & per cent \\
        \hline
        0.5 & 0.5 & 1 & 60.8\\
        1 & 1 & 1 & 73.2\\
        3 & 3 & 1 & 77.9\\
        5 & 5 & 1 & 78.3\\
        1 & 5 & 1 & 78.2\\
        5 & 1 & 1 & 76.1\\
        3 & 3 & 3 & 78.1\\
        5 & 5 & 5 & 78.4\\
        \hline
    \end{tabular}
\end{table}

The other main numerical parameter is the time interval between successive checks. Each of our N-body simulations outputs 1000 equally spaced snapshots, where the first snapshot is at time zero, and the final snapshot is at \mbox{10 diffusion} timescales. \mbox{Table \ref{tab:stirring_intervals}} shows how the measured stirring level for the projectile simulation on \mbox{Figure \ref{fig: projectileVsPltOnlySims}} changes as a function of the interval between checked snapshots. Based on these results, we set the numerical stirring analysis to check every 20 snapshots.

\begin{table}
    \centering
    \caption{Variation of the measured stirring level for the projectile simulation on Figure \ref{fig: projectileVsPltOnlySims}, as a function of the interval between checked {\sc rebound} snapshots. In these examples the parameters PNR, PIR and IR were set to ${1\percent}$, ${1\percent}$ and ${1\degree}$ respectively. Particles were considered scattered if their semimajor axes changed by more than ${20\percent}$. The values shown are the fraction of debris that is stirred but not scattered.}
    \label{tab:stirring_intervals}
    \begin{tabular}{cc} % two columns, alignment for each
	  \hline
	  Interval between checked snapshots & Fraction of disk stirred / per cent \\
        \hline
	  1 & 74.0\\
        2 & 73.9\\
        3 & 73.4\\
        4 & 73.4\\
        5 & 73.2\\
        10 & 73.2\\
        20 & 73.0\\
        50 & 72.0\\
        100 & 71.4\\
        \hline
    \end{tabular}
\end{table}

Using these analyses, the code classifies each debris body as either stirred, scattered or unstirred. A stirred particle is defined as one that satisfies the above conditions, and a scattered body is defined as one whose semimajor axis changes by more than ${20\percent}$ throughout the simulation; all other debris bodies are classed as unstirred.

%#############################################################################################################################
\section{Resonant-stirring efficiency versus other system parameters}

\begin{figure*}
    % projectileAndResonantStirring/images/stirringIncreaseByProjectiles/stirringIncreaseByProjectiles.vsz
    \centering
    \includegraphics[width=17cm]{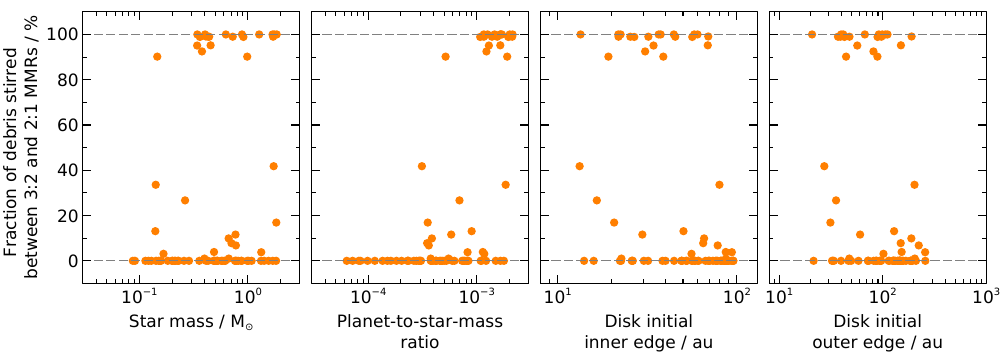}
    \caption{Resonant-stirring efficiency as a function of several system parameters. Figure \ref{fig: resonantStirringVsPlanetMass} showed that planet mass appears to be clearly linked to resonant-stirring efficiency; conversely, the remaining parameters shown here are less directly linked to whether resonant stirring occurs.}
    \label{fig: resStirringAppGrid}
\end{figure*}

Figure \ref{fig: resStirringAppGrid} shows the efficiency of resonant stirring as a function of several system parameters.

%#############################################################################################################################

% Don't change these lines
\bsp	% typesetting comment
\label{lastpage}
\end{document}